# USING NARROW BAND PHOTOMETRY TO DETECT YOUNG BROWN DWARFS IN IC348

A. K. Mainzer[1] and Ian S. McLean[1]
[1]Department of Physics and Astronomy, University of California, Los Angeles, CA 90095; mainzer@astro.ucla.edu; mclean@astro.ucla.edu

## ABSTRACT

We report the discovery of a population of young brown dwarf candidates in the open star cluster IC348 and the development of a new spectroscopic classification technique using narrow band photometry. Observations were made using FLITECAM, the First Light Camera for SOFIA, at the 3-m Shane Telescope at Lick Observatory. FLITECAM is a new 1–5 μm camera with an 8 arcmin field of view. Custom narrow band filters were developed to detect absorption features of water vapor (at 1.495 μm) and methane (at 1.66 μm) characteristic of brown dwarfs. These filters enable spectral classification of stars and brown dwarfs without spectroscopy. FLITECAM's narrow and broadband photometry was verified by examining the color-color and color-magnitude characteristics of stars whose spectral type and reddening was known from previous surveys. Using our narrow band filter photometry method, it was possible to identify an object measured with a signal-to-noise ratio of 20 or better to within ±3 spectral class subtypes for late-type stars. With this technique, very deep images of the central region of IC348 (H ~ 20.0) have identified 18 sources as possible L or T dwarf candidates. Out of these 18, we expect that between 3 – 6 of these objects are statistically likely to be background stars, and the remainder being true low-mass members of the cluster. If confirmed as cluster members then these are very low-mass objects (~5 $M_{Jupiter}$). We also describe how two additional narrow band filters can improve the contrast between M, L, and T dwarfs as well as provide a means to determine the reddening of an individual object.

## 1. INTRODUCTION

Prior to the discovery of brown dwarfs (Becklin & Zuckerman 1988, Oppenheimer et. al. 1995), the main stellar sequence ended with spectral class M. Stars of spectral class M are characterized by the appearance of deep molecular absorption bands due to the formation of oxides of refractive elements, such as Titanium Oxide (TiO) and Vanadium Oxide (VO). Recent years have seen the extension of the spectral sequence to lower temperatures and substellar masses with two new classes, L (with temperatures between 2500 and 1500 K) and T (with temperatures less than 1500 K). L dwarfs are characterized by the absence of TiO and VO absorption and the presence of metal hydrides and strong alkali lines in their optical spectra, and by strong water vapor absorption bands in the J, H, and K infrared wavebands (e.g. Martin et al. 1999; Kirkpatrick et al. 1999, 2000). T dwarfs are distinguished by the strengthening of water absorption bands and the onset of methane absorption in J, H, and K (e.g. Burgasser et al. 1999, 2000; Geballe et al. 2002). Until the discovery of the T dwarf Gliese 229B (Nakajima et al. 1995, Oppenheimer et al. 1995) methane absorption had only been seen in the spectra of the Jovian planets and Titan. Spectroscopic surveys, such as the NIRSPEC Brown Dwarf Spectroscopic Survey (BDSS) at the Keck Observatory (McLean et al. 2003) and others (Geballe et al. 2002), have begun to systematically characterize M, L, and T dwarfs in the infrared and produce canonical, high signal-to-noise spectra of each subtype. Figure 1 shows a number of NIRSPEC spectra from the BDSS. The transitions between M, L, and T dwarfs can be seen. The appearance (and strengthening) of water absorption at 1.5 μm characterizes the L dwarfs, and the addition of methane at 1.66 μm distinguishes the T dwarfs. Theory suggests that L dwarfs are a mixture of hydrogen-burning stars and sub-stellar brown dwarfs. All T dwarfs have sub-stellar masses.

Most of the brown dwarfs discovered to date in large-area sky surveys (2MASS, DENIS, SDSS) are located relatively close to the Sun, typically within 30 pc (Dahn et al. 2002), and are intrinsically very faint. In the last few years several investigators have searched for more distant brown dwarfs by studying young star clusters where any low-mass objects would be more luminous at this stage of their evolution (Zapatero Osorio et al. 2002; Najita et al. 2000; Lucas & Roche 2000). Surveying young clusters has several advantages. A large sample with a similar age is obtained, since the stars are spatially grouped together and tend to be roughly coeval. In contrast, the ages of field objects are not as well-known. Several models of young brown dwarfs have been made to predict their luminosity and colors as a function of age and mass (see Burrows et al. 2001 for a review). Since brown dwarfs have no sustainable internal fusion energy source, their temperatures steadily decline with time. Sub-stellar mass objects are two to five orders of magnitude brighter when young , making it possible to detect much lower-mass objects at greater distances. Young clusters also have smaller physical sizes than older optical clusters, lessening the effect of contamination by foreground field brown dwarfs.

The cluster we have chosen (IC348) is at -20° galactic latitude, sufficiently out of the galactic plane to lessen the degree of field star contamination. However, it can still be difficult to establish membership in the cluster and difficult to establish the spectral class of the object on the basis of broadband colors alone. We therefore sought to develop a photometric technique that would more readily allow identification of spectral type. The primary scientific goals of this project were three fold:

1) Develop a set of custom narrow band indices useful for spectral classification by photometric imaging without the need for spectroscopy. Indeed, for the faintest objects, it may not be possible to obtain even low-resolution spectra, therefore demanding alternate spectral classification methods. The goal was to develop a set of indices that excel at separating out the lowest mass objects, late M, L and T dwarfs, from every other spectral type. We created two custom narrow band filters suitable for detecting strong, broad spectral features unique to very low-mass stars and brown dwarfs. The selected bands were water ($H_2O$) at 1.495 μm and methane ($CH_4$) at 1.66 μm. Both features are in the H band, which was chosen for high detection sensitivity and minimal attenuation due to extinction within the cluster.

2) Use these narrow band filters to survey a young star cluster for the lowest mass members, down to planetary mass objects (5 $M_{Jupiter}$), over a wide enough area of the cluster to obtain meaningful statistics.

3) Prove, through the use of standard stars and follow-up spectroscopy that the narrow band classification scheme works reliably to identify an object's spectral type.

IC348 is 300 pc away (Herbig 1998) with an age of ~3 Myr (Luhman et al. 1998). Figure 2 shows the predicted change in magnitude for substellar objects of various masses and ages at a distance of 300 pc based on recent models (Burrows et al. 2001). From this figure, it can be seen that any cluster members detected at 19$^{th}$ magnitude should be between 2–10 $M_{Jupiter}$. Figure 3 shows the predicted change in temperature as a function of mass and age (also from Burrows). From this figure, we see that at 3 Myr, an object which is cool enough to exhibit methane absorption would be between 1 – 5 $M_{Jupiter}$. Other surveys for brown dwarfs in IC348 have been made, including Muench et al. (2003: hereafter M2003), Luhman (1999: L99), Luhman et al. (1998: L98), Lada & Lada (1995), and Najita, Tiede, & Carr (2000: N2000). Most previous studies have been limited to $m_H$ = 18. With FLITECAM, a new infrared camera built at UCLA (Mainzer et al. 2003), we have surveyed down to $m_H$ = 20.0 with SNR = 3 over a ~18 arcmin$^2$ region.

## 2. THE OBSERVATIONAL TECHNIQUE

### 2.1 DEVELOPMENT OF THE PHOTOMETRIC INDICES

To determine the best set of narrow band indices to use for spectral classification, a model was created based on the flux-calibrated spectra of 85 objects for which R~2000 spectra had already been obtained. These objects ranged in spectral type from B0 main sequence stars to M9.5 dwarfs, early and late L and T dwarfs, M and K giants and supergiants, and galaxies. The main sequence stars were drawn from a database created by Pickles (1998); the M dwarfs from Leggett et al. (2002), the galaxy spectra from M. Rieke (private communication), and the late M, L and T dwarf spectra from the BDSS.

Using the desired set of trial filter bandpasses as input, the model calculates the magnitudes of each object in each band by numerically integrating the total flux in that band. In addition, the model creates color-magnitude and color-color plots for each set of filter combinations. The primary characteristics of T dwarfs are strong $H_2O$ and $CH_4$ absorption bands throughout the near-infrared (see Figure 1). Therefore, a set of indices based on the strengths of water and methane absorption relative to a continuum band should yield the best discriminator of these objects.

Although strong $H_2O$ and $CH_4$ features are also found in the K band, it was determined that the best choice was the H band using features centered at 1.495 μm and at 1.66 μm. For ground-based observations, the K band is less sensitive due to thermal background emission from the telescope. Although the J band could be used for the continuum filter, the lower thermal background at J band is offset by high extinction by the dust shrouding the young stars in the cluster. Both J and H suffer from variable OH emission in the Earth's atmosphere, but H provides the best overall sensitivity.

By examining the spectra of L and T dwarfs from the BDSS, it became obvious that the optimal filters for detecting water and methane should have widths of 0.1 μm, a 6% width for both filters. The goal was to widen the bandpass as much as possible to allow maximum throughput in that absorption feature while maintaining sensitivity to the spectral structure. Figure 1 shows the BDSS spectra of several M, L and T dwarfs with the narrow-band filters overlaid. Initially, a third narrow band filter centered on 1.595 μm had been planned. This filter would have sampled the continuum between the water and methane features and would have provided the best contrast between methane-bearing T dwarfs and everything else. However, the additional cost for the third filter was prohibitively high at the time and so we postponed its purchase, substituting the H filter as a proxy for the continuum. Figures 4 and 5, created from the model, compare color-color plots using the broadband H filter and using the narrow continuum filter. As shown, the continuum filter more distinctly separates out the T dwarfs from the rest of the stars and galaxies, providing greater contrast, but the H band provides a reasonable approximation for an initial survey to test the technique. A narrow H continuum filter has now been purchased and further observing runs will be carried out using the new filter. Figure 6 shows the as-built $H_2O$ and $CH_4$ filter transmission curves. These filters differ from the narrowband NICMOS filters F166N, F190N, and F215N used by N2000 in that the NICMOS filters are 1% wide instead of 6%. N2000 used the F166N filter to sample the $CH_4$ band, F190N to sample $H_2O$, and F215N as a proxy for continuum.

In IC348, typical reddening values range from an $A_V = 5$ to 10 (M2003, L99, L98). From the Rieke and Lebofsky (1985) interstellar extinction law, this translates to $A_H \sim 1 - 2$. This reddening law was applied to all the spectra in the bandpass model, allowing us to compare different color-color plots to see which ones yield useful information in the presence of reddening. It was immediately clear that Figure 4, which plots H - [1.495] versus H – [1.66], does the best job at separating out the methane-rich T dwarfs, particularly when reddening is considered. In this figure, stars progress right to left from the main sequence (shown in blue) through the M (red crosses) and L (green crosses) spectral classes along a mostly horizontal line. The T dwarf sequence (shown in cyan) progresses from the end of the L dwarfs in a line almost perpendicular to the main sequence to the lower left. This happens because methane absorption at 1.66 μm increases in strength through the sequence. Other color-color plots, including the standard H – K versus J – H plot (shown in Figure 7) are not as helpful in discriminating T dwarfs from stars within clusters when reddening is considered. One color-color plot that did appear useful as a discriminator is the z – [1.495] versus z – [1.595] plot since the T dwarf sequence is perpendicular to the reddening vector (Figure 8). However, the z band suffers from much more extinction than H so many faint sources are simply undetectable. Nevertheless, since z band is a potentially useful discriminator, we commissioned a custom z band filter. The half-power points of our z filter are at 0.95 μm and 1.10 μm. Luhman et al. 2003 have recently completed I and z band surveys of IC348; however, due to extinction, their survey is not quite deep enough to reliably detect T dwarfs using Figure 8.

## 2.2 SPECTROSCOPY OF AN M8 DWARF IN IC348

During a BDSS run at the Keck Observatory in late August, 2002, we selected a candidate low mass object from the L98 survey to observe with NIRSPEC. We chose object #611 (hereafter LRLL611), which is classified in L98 as an M8 with zero reddening. In IC348, the hydrogen-burning limit is equivalent to an M6 star (L98), therefore at M8, LRLL611 is probably sub-stellar. Figure 9 shows its H band spectrum (from the NIRSPEC N5 filter). Water absorption at 1.5 μm appears ~30% deeper in the young cluster member LRLL611 than it does in the more evolved field M8 dwarfs. This result is most likely due to lower surface gravity in the young M8 compared to the old field dwarf counterpart. Similar behavior can be observed with late-type field M giants. Of course, the increased water absorption could also be explained if the object's spectral type is actually later than that measured by L98, or if its reddening is not actually zero.

Once the digitized, flux-calibrated spectrum of LRLL611 was obtained, it could be placed with the other spectra in the model. LRLL611 has $A_H$ = 0 (L99). Its lack of reddening makes it a good candidate to calibrate the model to see where an unreddened young M8 falls. When it was added to the bandpass model (see the black cross in Figure 4), LRLL611 fell into place in the color-color plot among the other late M and L dwarfs. However, since the NIRSPEC N5 filter does not completely sample the bandpasses equivalent to the new [1.495] filter and the broadband H filter, pieces had to be extrapolated onto the ends of the spectrum. Depending on the extrapolation used, LRLL611's synthesized H – [1.495] value ranged between -0.54 and -0.79, with the more negative value corresponding to a deeper water absorption feature. Overall, however, this result increased our confidence prior to the start of FLITECAM observations that the narrow-band color-color spectral typing would be successful. Future spectroscopic observations will cover the entire bandpass.

## 2.3 THE EFFECT OF REDDENING

As can be seen in Figure 4, the reddening vector as computed from the Rieke-Lebofsky reddening law is perpendicular to the T dwarf series, but unfortunately it is parallel to the line containing the main sequence stars, M dwarfs, L dwarfs, and T1 and T2 dwarfs. Reddening moves objects ~ -0.3 magnitudes in the H – [1.495] axis for $A_V$ = 10 and a negligible amount (~0.01 magnitudes) in the H – [1.66] axis. Hence, objects in this region of the color-color plot are degenerate with reddening. It is impossible to tell which type they are unless the reddening is known from some other means.

Figure 10 depicts the spectra of several objects whose classification according to the H - [1.495] vs. H - [1.66] color-color plot is indeterminate when variable reddening is included. These are reddened M dwarfs, K and M giants, L dwarfs, and T1 and T2 dwarfs. As shown in the figure, an M8 giant has a much deeper water bands than does a field M8 dwarf. Deeper absorption bands are the result of the lower surface gravity in the giant star. Thus, attributing LRLL611's deeper water band to an extended atmosphere due to its youth seems a plausible explanation.

### 2.4 BACKGROUND OBJECTS

The greatest difficulty when using deep observations of a cluster to spectrally classify its members is distinguishing genuine members from background and foreground objects. One method of tackling this problem is to observe an off-cluster region at a similar galactic latitude/longitude to the same or greater depth as the cluster. An estimated cluster reddening can be applied to these background stars to derive the background counts likely to have been seen in the cluster. The disadvantage of this method is that it provides statistical information only. It does not identify which individual objects are likely to be cluster members versus background stars. The method described in this paper uses narrow band filters to spectrally classify individual objects based on their narrow band colors, avoiding much of the ambiguity.

## 3. OBSERVATIONS AND RESULTS

### 3.1 INSTRUMENT

All our observations of IC348 were performed with FLITECAM (Mainzer et al. 2003), the First Light Camera for the Stratospheric Observatory for Infrared Astronomy (SOFIA). FLITECAM is a facility-class cryogenic instrument being developed at UCLA (Principal Investigator: I. S. McLean). FLITECAM performs imaging between 0.9 and 5 μm over the full unvignetted 8' field of view of the SOFIA telescope. Since it is a "first light" instrument, we began verification of FLITECAM well in advance of its 2004 delivery date to SOFIA by deploying it at the University of California's 3-m Shane Telescope at Lick Observatory.

FLITECAM's imaging scale is ~0.43 arcsec per pixel using a 1024x1024 indium antimonide (InSb) ALADDIN array operated at 30 K. Large refractive optics cooled to ~77 K are used for collimation and re-imaging, and the instrument includes standard J, H, K, K', L, L', and M passband filters, as well as a limited set of narrow band filters. Later, FLITECAM will also be equipped with a pupil-viewing mode optimized for 3.5 μm and with grisms in its collimated beam for long-slit R~2,000 spectroscopy.

Commissioning began at Lick Observatory on 2002 September 13. The Shane Telescope was selected because the telescope's F/17 focal ratio is nearly compatible with the F/19.6 SOFIA optical system for which FLITECAM was designed. Two additional observing runs followed in October and November 2002 during which IC348 was observed. Table 1 summarizes the general instrument characteristics of FLITECAM.

| Characteristic | Value |
|---|---|
| Wavelength range | 0.9 to 5.5 μm; L band for pupil viewing |
| Filters | J, H, K, K', L, L', M, plus narrow band |
| Spectral resolution | 1200 – 2000 in Grism mode |
| Spatial resolution | 0.43" per pixel |
| Detector type | InSb Raytheon ALADDIN III |
| Detector format | 1024x1024 |
| Field of view | ~8 arcmin diameter |
| Detector operational temperature | 30 K |
| Cryostat type | 20 L liquid nitrogen / 20 L liquid helium |
| Read noise | ~50 electrons |
| Well depth | ~80,000 electrons |
| Dark current | ≤ 1 electron/sec |
| Sensitivity | SNR=20, ~20 minutes, H=18.0 |
| Instrument efficiency | ~40% (not including QE) |
| Detector quantum efficiency | ~80% |
| Typical seeing | ~1.1 – 1.4 arcsec |

Table 1: Summary of the general characteristics of FLITECAM on the Lick 3 m Shane Telescope.

3.2 PHOTOMETRY

To accurately distinguish bona fide sub-stellar cluster members from background objects, models showed that a 5% photometric accuracy, or a signal-to-noise ratio (SNR) = 20 was required. In addition, to provide a reasonable chance of finding young methane-bearing objects in IC348, our survey needed to be to a minimum depth of 18$^{th}$ magnitude in H, [1.66], and [1.495] with that SNR. We achieved this goal, surveying to a depth of $m_H$ = 18.5 with SNR = 20 in all filters. Since other surveys have covered larger areas to shallower limits, we picked a single field near the center of IC348 and went as deep as possible. The observations reported here were obtained from two runs, 2002 October 29-31, and 2002 November 11-13. The total integration times were 4.5 hours in $H_2O$, 4.5 hours in $CH_4$, and 1.1 hours in H. The total survey area was 18 square arcminutes. This is smaller than the nominal field of view due to a failure of half the detector array three days prior to the start of the runs caused by a faulty cable internal to the cryostat; the fault has since been corrected.

The FLITECAM images were dark subtracted, flat fielded, and source extracted using a custom-written software package optimized for crowded and nebulous regions. FLITECAM H band photometry agreed with 2MASS and the photometry of L98 and L99 to within ±0.1 magnitudes on 45 objects fainter than $m_H$ = 11.6 (FLITECAM data begin to saturate at brighter magnitudes) and brighter than $m_H$ = 15.7 (at which point 2MASS becomes incomplete and noisy). FLITECAM H band photometry agreed with the recently published FLAMINGOS H band photometry to within ±0.1 magnitudes on 45 objects with magnitudes between 10 – 16, and to within ±0.4 magnitudes on ~50 objects with magnitudes between 16 – 18 (M2003).

The photometric zero points for the narrow band water and methane filters were determined by examining a star known from Luhman's spectroscopic measurements to be an A2. An A2 star should have neither water nor methane absorption, and hence should have almost neutral colors. The ratio of [1.495] to [1.66] to H should be 1:1:1. The zero points of the narrow band filters were adjusted such that the A2 star had the same magnitude in all three bands. The on-chip integration time for the narrow bands was boosted by a factor of three relative to the broadband H since the narrow band filters are one-third the width of the H filter.

With such a large field of view, the FLITECAM point spread function varies across the field. However, since the three bands of interest ([1.495], [1.66], and H) are all extremely close in wavelength, relative photometry between bands remains accurate since the same amount of optical distortion occurs in all three bands. Absolute photometry of objects in the edge regions of the field is degraded since flux is scattered outside the aperture radius.

### 3.3 SIMULATED STARS

To verify photometric accuracy, several hundred simulated star locations were created to enable the insertion of artificial stars into the H, [1.495], and [1.66] images. The same custom-written photometry routine was run on the simulated stars as on the real stars. The simulated stars were scattered randomly throughout the images in all the areas where real objects were detected. To accurately model FLITECAM's detection statistics, simulated stars were inserted with a $dN/dm_H$ profile drawn directly from the real stars measured in IC348 between $m_H$ = 16–19.5. A power law was extrapolated from the observed $dN/dM_H$ behavior over these magnitudes down to $m_H$ = 21. In this way, faint stars that were scattered by random noise to brighter magnitude bins were more accurately modeled.

### 3.4 SURVEY LIMITS

The simulated stars allowed an estimate of the signal-to-noise ratio (SNR) versus magnitude to be made. An average sensitivity can be estimated by plotting magnitude vs. SNR using the real and simulated stars. The mean 3-$\sigma$ detection limit was 20.0 for H, [1.495], and [1.66].

### 3.5 VERIFICATION OF NARROW BAND CLASSIFICATION TECHNIQUE

Figure 11 shows the color-color plot resulting from FLITECAM measurements of 45 stars for which spectral classes have been independently measured spectroscopically by L98 and L99. Spectral classes for these objects range from A2 through M8.25. They are color-coded by spectral type. Many of these objects had estimates of their reddening determined by L98. The objects represented here have been dereddened by this amount to move them to their unreddened locations in the color-color plot. In Figure 11, not all the sources have measured reddening, so some of them appear farther to the left than they should be. As shown in the figure, the main sequence can be clearly traced from A2 through M8.25.

Figure 12 shows the model predictions for this same color-color plot without reddening. According to the model, the objects at the end of the sequence to the left can either be reddened M dwarfs, unreddened and reddened L and T dwarfs, and possibly an M9 giant. With this filter set, an $A_V$ of 10 will move objects 0.3 magnitudes to the left in the H - [1.495] axis only; this amount of reddening results in a negligible (~0.01 magnitude) motion in the H - [1.66] axis. This independent calibration method using IC348 itself validates the narrow band photometric identification technique for stars with spectral classes A-M. Efforts were made to observe a series of known field L and T dwarfs with FLITECAM in all bands. However, non-photometric conditions meant that it was impossible to place the standards onto the color-color plot with the IC348 objects to sufficient accuracy. This program to observe L and T standards will be carried out in future runs.

A set of color-color indices can be derived from these observations of previously classified young stars and brown dwarfs in IC348 and compared to similar indices derived from the bandpass model described in Section 2. Table 2 lists the H – [1.495] and H – [1.66] indices derived for the various spectral classes from the integrated spectra bandpass model. Table 3 lists the same indices derived for the dereddened IC348 cluster members. We found a trend toward more negative H – [1.495] colors in the dereddened IC348 objects compared to the field objects (see Figures 11 and 12). Typically, the young stars and brown dwarfs showed about 0.2 – 0.3 magnitudes more negative H – [1.495] color compared to the bandpass model objects. This difference can possibly be explained by the differing surface gravities of the two populations, since the bandpass model was constructed using spectra from field objects, most of which are >1 Gyr old, compared to the ~3 Myr age of the IC348 objects. It is also of interest to note that the field objects' H – [1.66] values typically increase ~0.2 magnitudes when progressing from early to later spectral types. However, the IC348 objects have H – [1.66] colors that trend in general in the opposite sense, tending to decrease ~0.1 magnitudes. Perhaps this effect as well could be explained by differing surface gravity. Future experiments will examine the variation of these indices with age and surface gravity.

A large part of the uncertainty in our ability to classify the IC348 objects was due to the uncertainties in the reddenings derived by L98 and L99 – any error in these values contributes directly to classification error. If three stars for which the reddenings listed in L98/L99 were either not known or appear to be underestimated (sources LRLL232, LRLL248, and LRLL221) are not used for deriving the color indices, we obtain the results shown in Table 3. With this filter set, we can see from Tables 2 and 3 that it appears possible to classify bright stars to within approximately ±3 spectral class subtypes for late-type stars if the individual objects' reddening values are known from another source. A filter set that allows for measurement of a reddening-independent index will be employed in future observing runs.

| Field Objects' Spectral Classes | H – [1.495] range | H – [1.66] range |
|---|---|---|
| B | -0.12 to 0.00 | 0.00 to 0.03 |
| A | -0.18 to -0.15 | 0.03 to 0.04 |
| F | -0.15 to -0.11 | 0.03 |
| G | -0.19 to -0.13 | 0.03 to 0.04 |
| K | -0.30 to -0.20 | 0.05 to 0.07 |
| M1-M3 | -0.31 to -0.24 | 0.05 to 0.08 |
| M4 – M5 | -0.40 to -0.27 | 0.07 to 0.10 |
| M6 – M7 | -0.33 to -0.46 | 0.08 to 0.12 |
| M8 | -0.42 to -0.47 | 0.09 to 0.12 |
| M9 | -0.55 to -0.45 | 0.09 to 0.13 |
| L0 | -0.54 to -0.49 | 0.12 to 0.13 |
| L2 | -0.82 to -0.56 | 0.11 to 0.20 |
| L3 – L4 | -0.68 to -0.72 | 0.14 to 0.17 |
| L5 – L7 | -0.77 to -0.71 | 0.16 to 0.17 |
| L8 | -0.76 to -0.74 | 0.17 to 0.18 |
| T1 – T2 | -0.90 to -0.86 | 0.20 to 0.22 |
| T5 | -0.78 to -0.81 | -0.05 to 0.02 |
| T6 | -0.78 to -0.66 | -0.24 |
| T7 | -0.52 | -0.50 |
| T8 | -0.77 | -0.65 |

Table 2: H – [1.495] and H – [1.66] indices derived for field objects (using the bandpass model and spectra).

It also appears from Tables 2 and 3 that the H – [1.495] index decreases more significantly in the later spectral types (M6 – M8.25) in the young IC348 objects than in earlier spectral classes. The effects of surface gravity on the H – [1.495] index appear to be slightly more pronounced in later spectral types; the difference between field object and IC348 H – [1.495] indices for earlier spectral classes is ~0.1 magnitudes, compared with ~0.25 for M dwarfs. Lucas et al. 2001 and Gorlova et al. 2003 provide further discussion of the effects of surface gravity and age at these wavelengths.

| IC348 Objects' Spectral Classes (from L98/L99) | H – [1.495] range | H – [1.66] range |
|---|---|---|
| A2 – G7 | -0.03 to 0.00 | 0.00 to 0.02 |
| K2 – K8 | -0.40 to -0.02 | -0.04 to 0.03 |
| M0 – M1 | -0.45 to -0.30 | -0.07 to 0.00 |
| M2 – M3 | -0.58 to -0.23 | -0.11 to 0.03 |
| M4.5 – M5 | -0.59 to -0.49 | -0.12 to -0.02 |
| M5.75 – M6.25 | -0.65 to -0.42 | -0.06 to 0.00 |
| M8 – M8.25 | -0.79 to -0.70 | -0.07 to -0.04 |

Table 3: H – [1.495] index ranges for 3 Myr old IC348 objects. Note that colors for IC348 objects have been dereddened using reddening estimates from L98/L99.

### 3.6 IDENTIFICATION OF BROWN DWARF CANDIDATES

Figure 13 shows all the stars measured by FLITECAM, with the 45 stars whose spectral classes were previously known shown in the colors of Figure 11. Figure 14 shows all stars measured by FLITECAM (shown as colored circles), plus the results of a model involving 25,000 simulated stars (contours). The contours represent only simulated stars with $m_H$ = 18 – 19.5. These simulated stars all had input colors consistent with M0 – M3 stars with a reddening equivalent to $A_V$ = 10. In this way, we test the null hypothesis that there are no T dwarfs, and that all faint objects are early-type M and K background stars seen through $A_V$ = 10 reddening. The contours represent the regions in which 50%, 68%, and 92% of the simulated stars with $m_H$ = 18 – 19.5 are contained. It appears that fainter real stars (for which the spectral types have not been previously measured) are preferentially scattered to the lower left of the main sequence where, according to the model, objects with both water and methane absorption are located. However, we must rigorously test the camera's detection statistics to determine whether or not this is a real effect. If faint objects were actually all main sequence M dwarfs and K giants, one might suspect that they would be evenly distributed on either side of the main sequence. Since reddening can only move objects to the left, no other process other than methane absorption or statistical error can move objects to the lower left of the diagram. The reason the photometric errors as demonstrated by the simulated stars appear in a tilted ellipse rather than a Gaussian aligned with the axes is that, since the H band has been used in both axes, the errors in the two axes are correlated. Thus, scattering either faint or bright in H moves a star in both axes. A crude zeroth-order analysis can be made by simply considering the number of observed stars with magnitudes between 18 – 19.5 that are above and below the zero boundary of H – [1.66]. According to the model, to first order there should be no methane absorption if the objects are not T dwarfs. We count that 49 out of 72 objects with $m_H$ = 18 – 19.5 show H – [1.66] < 0.

However, a much better estimate of the statistical noise can be made by considering where objects fall relative to the simulated star contour boundaries. We counted the number of both simulated and real objects that were inside a region bounded by the limits $-1.5 < (H - [1.495]) < 0$ and $-0.5 < (H - [1.66]) < 0$, as well as the 92% contour boundary (the "brown dwarf box"). These are the regions in which brown dwarf/planetary mass objects would be found according to the bandpass model. We expected to find 3.1 real stars in this region and found 18. We then considered the region bounded by the 92% contour boundary but outside the brown dwarf box. Objects in this region should only be present due to statistical scattering, according to the bandpass model. In these regions of the color-color plot, we expected to find 2.5 real stars, and we found 5. This could be indicative of a population of water and methane-bearing objects, but these statistics must be viewed with caution since they were derived from a simulated background population.

Thus, we do not appear to have detected any large population of methane-bearing objects. However, we detected 18 objects with $m_H = 18 - 19.5$ that are in the brown dwarf box but outside the 92% contours traced by the simulated stars (see Figure 14). Out of these 18, we expect that 3 - 6 are background objects. Depending on their location in the brown dwarf box, these could either be: 1) reddened late-type M stars; 2) extremely reddened ($A_V > 12$) early M stars; 3) reddened or unreddened L dwarfs of various types; or 4) T dwarfs. At an age of 3 Myr and a distance of 300 pc, a T dwarf (defined as an object with a temperature less than ~1500 K, and therefore cool enough to exhibit methane absorption) with $m_H = 19$ would be ~1-5 $M_{jupiter}$. (Burrows et al. 2001) (see Figure 4). Table 4 lists the objects observed by FLITECAM.

Additional statistical techniques may be employed to determine whether or not the detected object population differs from the simulated stars. A Kolmogorov-Smirnov test indicates that the observed stars' H - [1.66] magnitudes differ by 17% from the simulated stars, with a 0.05% chance of coming from the same population. The H - [1.495] magnitudes differ by 22% with a 3% chance of coming from the same population. However, these tests must be viewed with caution since, a) the Kolmogorov-Smirnov test is a one-dimensional test capable of considering only the difference between two colors, not three; and b) the K-S test only compares the difference between the observed stars and the simulated stars, which have been modeled to have the colors of reddened early M stars. The colors of the true background have not been measured, since we did not make any off-cluster measurements due to time and weather constraints. An off-cluster measurement will be made in future observing runs. Once a background measurement has been made, an estimated reddening can be applied to the background population.

From the broadband H photometry, it is possible to construct a cluster luminosity function. Figure 15 shows $d\log N/dm_H$. A bimodal distribution is apparent and confirms the results of L98, L99, N2000, and M2003. The first peak appears at $m_H \sim 12$ and tapers off by $m_H \sim 16$. The second peak, which consists primarily of background objects, begins here and rises with a power law distribution with a slope of 0.24 up to the completeness limit of 19.5$^{th}$ magnitude.

The number of background stars and background galaxies that should be in the FLITECAM field of view can be estimated. An estimate of the number of galaxies (Madau & Pozzetti 2000) yields ~9 unreddened background galaxies per FLITECAM field of view. M2003 surveyed IC348 to a completeness limit of $m_H$ = 18. From that work we find that numbers of background stars do not become significant in comparison to cluster members until $m_H$ ~ 17. Extrapolating their number counts down to $m_H$ = 19.5 resulted in a prediction of more stars than FLITECAM observed at fainter magnitudes, although agreement is good at brighter magnitudes. This could be due to the fact that by $m_H$ = 19, star counts drop as we see beyond the edge of the galactic disk population at this galactic latitude (although total background objects will not drop due to confusion with unresolved galaxies). A typical M3 dwarf with $m_H$ = 20.5 will be visible out to ~1500 pc and is beyond the edge of the galactic disk, which is ~200 pc away (Bahcall & Soneira, 1988). Thus, background counts from the M2003 observations cannot be used to accurately extrapolate background object counts at faint magnitudes to obtain an estimate of number of cluster members. We are planning future off-cluster observations to assess the background contamination.

It is possible to make a rough estimate of foreground contamination by using estimates of the density of field T dwarfs, which is approximately one per 160 $pc^{-3}$ (Collinge et al. 2003). We surveyed 18 $arcmin^2$ with FLITECAM down to $m_H$ = 20 3-$\sigma$; therefore, we anticipate between 0.002 – 0.007 foreground T dwarfs in our field of view (assuming we can detect a T dwarf out to ~75 pc). Thus, foreground T dwarfs are not likely to be a significant contributor to our source counts.

This analysis does not take into account the effects of binarity on the photometric colors. With 1” seeing at 300 pc, FLITECAM can resolve objects separated by > 300 AU. If an object that is cool enough to exhibit methane absorption is orbiting another hotter object with a separation < 300 AU, FLITECAM cannot distinguish the two, and their apparent photometric colors will be dominated by the brighter object. Since any newly discovered substellar objects are a small fraction of the total population, binarity is likely to result in an underestimate of the brown dwarf number counts. A discussion of the effects of binarity is given by M2003.

## 3.7 OTHER NARROW BAND FILTERS

The standard H filter encompasses both the water and methane absorption bands used in our filter set. Consequently, the H band continuum measurement is contaminated by the molecular absorption bands. The contrast between M, L, and T dwarfs can be significantly improved by using a custom narrow band filter for the continuum measurement instead of the broadband H filter. We have recently commissioned such a filter, centered at 1.595 μm and have begun using it. Figures 4 and 5 showed the improvement in contrast to be gained by using a narrow continuum filter instead of an H filter.

Potential gains can also be had by using four filters instead of three. In addition to the 1.495 μm water, continuum, and 1.66 μm methane filters, a second water filter optimized to pick up the short end of the 1.8 μm band should allow us to correct for reddening. The primary disadvantage of only using methane, water, and continuum filters is that the reddening vector remains parallel to the series of main sequence, M and L dwarfs, and early T dwarfs, rendering it impossible to tell a reddened M1 dwarf from an unreddened M6 or background star. The addition of the second longer wavelength water filter removes this degeneracy and allows us to solve for reddening. Instead of a two-dimensional color-color diagram, a three-dimensional color-color-color cube can be constructed (see Figure 16). The cube can be rotated to find the axis along which reddening is removed (see Figure 17). Reddened objects are plotted in red; unreddened objects are plotted in blue. This new 1.75 μm water filter has also been commissioned and will be employed in future FLITECAM observing runs.

### 3.8 CORRELATION WITH X-RAY DATA

Several surveys of IC348 have been made in recent years using the ROSAT and Chandra Observatories (Preibisch, Zinnecker, & Herbig 1996; Preibisch & Zinnecker 2001; Preibisch & Zinnecker 2002). L and T dwarfs are predicted to have little X-ray emission from their atmospheres, since their cooler temperatures cause them to be electrically neutral. However, in their Chandra survey of IC348, Preibisch and Zinnecker found that the photospheric properties of young substellar objects more closely resemble those of older, active stellar objects. Comparing the X-ray properties of young brown dwarfs with evolved field dwarfs could yield useful information about their levels of chromospheric activity versus age. Obtaining X-ray information could also help to more definitively identify cluster members, since X-ray emission generally will not come from distant old background stars. However, when the sources from Preisbisch and Zinnecker (2002) were correlated with IC348 IR targets, all X-ray sources but one correlated with known IR sources having H magnitudes brighter than 14, with the remaining source correlating with any IR target with an H magnitude of 15.5. No X-ray sources were associated with faint FLITECAM detections. A number of X-ray sources were found to have no correlation with any IR sources down to the FLITECAM limit, suggesting they are either background or originating from an undetected source.

## 4. CONCLUSIONS

We have described the development of a new technique using narrow band filters to detect the spectral class of very low mass objects without resorting to spectroscopy. We used a pair of custom narrow band filters optimized for detecting water vapor and methane, both characteristic signatures of very low mass objects. Using wide-field imaging through narrow band filters as a proxy for spectroscopy allows great potential savings in observing time.

Very deep imaging observations were made of the young cluster IC348 using these filters during commissioning of a new wide-field IR camera (FLITECAM). With the sensitivity to reach 18th magnitude at a signal-to-noise ratio (SNR) = 20 in ~20 minutes in broadband H, we were, in principle, able to detect objects with the mass limit of 2-5 $M_{jupiter}$ in IC348. Color-color diagrams were predicted beforehand using passband synthesis from existing spectra obtained either from the literature or from the NIRSPEC Brown Dwarf Spectroscopic Survey underway at UCLA. Data reduction techniques were developed to handle the crowded fields and large dynamic range in the IC348 observations.

For objects whose signal-to-noise ratios exceed 20, it is possible to determine spectral class photometrically to within ~±3 spectral class subtypes for late-type stars. Our technique was verified by observing 45 objects whose spectral classes were previously known through optical spectroscopy. We have created H – [1.495] and H – [1.66] indices for both old field objects and young cluster members. We have shown that there are ~18 objects between $m_H$ = 18 – 19.5 in the region of IC348 surveyed whose colors make them possible late M, L, or T dwarf candidates. We expect that between 3 – 6 of these objects are background sources due to noise. The number of background objects observed is consistent with the numbers predicted from a variety of observational and theoretical methods. Follow-up spectroscopic observations, although difficult, will allow verification of the methane and water spectral signatures.

The authors wish to thank the entire staff of the Infrared Lab at UCLA for their outstanding support in the development and deployment of FLITECAM. We also thank Eric Becklin, Adam Burrows, Michael Jura, James Larkin, and Erick Young for many useful comments on this work. Thanks are due to Jonathan Sievers for assistance with photometric reductions and to Mark McGovern for help with the NIRSPEC data. The authors also wish to thank an anonymous referee for many helpful suggestions. AKM was supported by a NASA Graduate Student Research Fellowship.

This research has made use of the NASA/IPAC Infrared Science Archive, which is operated by the Jet Propulsion Laboratory, California Institute of Technology, under contract with the National Aeronautics and Space Administration. FLAMINGOS was designed and constructed by the IR instrumentation group (PI: R. Elston) at the University of Florida, Department of Astronomy with support from NSF grant (AST97-31180) and Kitt Peak National Observatory. The data were collected under the NOAO Survey Program, "Towards a Complete Near-Infrared Spectroscopic Survey of Giant Molecular Clouds" (PI: E. Lada) and supported by NSF grants, AST97-3367 and AST02-02976 to the University of Florida.

Figure 1: A comparison of brown dwarf spectral classes, ranging from M6.5 through T8. The FLITECAM custom narrow band filters are superimposed. (Figure courtesy of Mark McGovern and the Brown Dwarf Spectroscopic Survey.)

Figure 2: The variation in broadband H apparent magnitude with age and mass for substellar objects at a distance of 300 pc. From Burrows et al. 2001.

Figure 3: The variation in temperature as a function of age and mass for substellar objects. From Burrows et al. 2001. At 3 Myr, any objects cool enough to exhibit methane absorption would be ~1 – 5 $M_{Jupiter}$.

Figure 4: Color-color plot using custom water and methane narrow-band filters, with a broadband H filter. Color code: blue = main sequence stars; red = M dwarfs; black = background galaxies; pink = K and M giants; green = L dwarfs; cyan = T dwarfs. The crosses represent a model of SNR = 20. The black arrow indicates the effect of 10 visual magnitudes of reddening.

Figure 5: Color-color plot using custom water, methane, and continuum filters. Color code same as Figure 3. The black arrow indicates the effect of 10 visual magnitudes of extinction.

Figure 6: FLITECAM's custom narrow-band water and methane filters.

Figure 7: Color-color plot using J, H, K filters. Color code same as Figure 3. The black arrow indicates the effect of 10 visual magnitudes of extinction. Reddening in a young cluster causes motion in color-color space, making it more difficult to discern spectral type on the basis of photometry alone.

Figure 8: Color-color plot using custom z, [1.595], and [1.495] filters. Color code same as Figure 3. The black arrow indicates the effect of 10 visual magnitudes of extinction. This color-color plot separates T dwarfs out from other objects well, but z band suffers heavily from extinction.

Figure 9: The H band spectrum of LRLL611, a known young (3 Myr) M8, observed with NIRSPEC on 1 September 2002. Since the NIRSPEC N5 filter was used, pieces at the short and long wavelength ends of the H band are missing.

Figure 10: A comparison of an M8 giant, an M8 field dwarf, an L5 field dwarf, and LRLL611, a young M8 dwarf. The 1.495 μm water feature is deeper in LRLL611 relative to field M8 dwarfs, similar to the M8 giant.

Figure 11: Color-color plot showing only stars with spectral types and reddenings known from L98 and L99. The stars are color-coded to reflect the main sequence, ranging from A2 through M8.25. The narrow band filter colors clearly trace out the main sequence,

allowing spectral classification to within ±3 spectral class subtypes for late-type stars without spectroscopy when reddening is known.

Figure 12: The model color-color plot without any reddening applied. Blue – A – K stars; red – M dwarfs; green – L dwarfs; pink – late K giants; cyan – T dwarfs; black – known M8 LRLL611. This model should be compared to the real data in Figure 10, for which spectral classes were known a priori.

Figure 13: Color-color plot showing FLITECAM IC348 targets. Color coding same as Figure 3, except for brown points, which are objects whose spectral types were not previously known.

Figure 14: Color-color plot showing both simulated stars (contours) between $m_H$ = 18 – 19.5, and all real stars detected by FLITECAM with $m_H$ < 19.5 (individual points). The contours represent the regions in which 50%, 68%, and 92% of the simulated stars were contained. The red circles represent real stars with $m_H$ = 9 - 15; yellow – real stars with $m_H$ = 15 - 18; teal – real stars with $m_H$ = 18 – 19.5.

Figure 15: Number of stars per 18 arcmin$^2$ FLITECAM field of view per half magnitude bin.

Figure 16: Color-color-color cube using two water filters at [1.495] and [1.75], a [1.66] methane filter, and a [1.595] continuum filter. The unreddened objects are shown in blue, beginning on the right with an A0 and progressing through the main sequence to the L dwarfs. T dwarfs project downward. Reddened objects are shown in red.

Figure 17: The same color-color cube as shown in Figure 16, but rotated to remove reddening. Now all spectral types, A through T, can be classified with no ambiguity due to reddening.

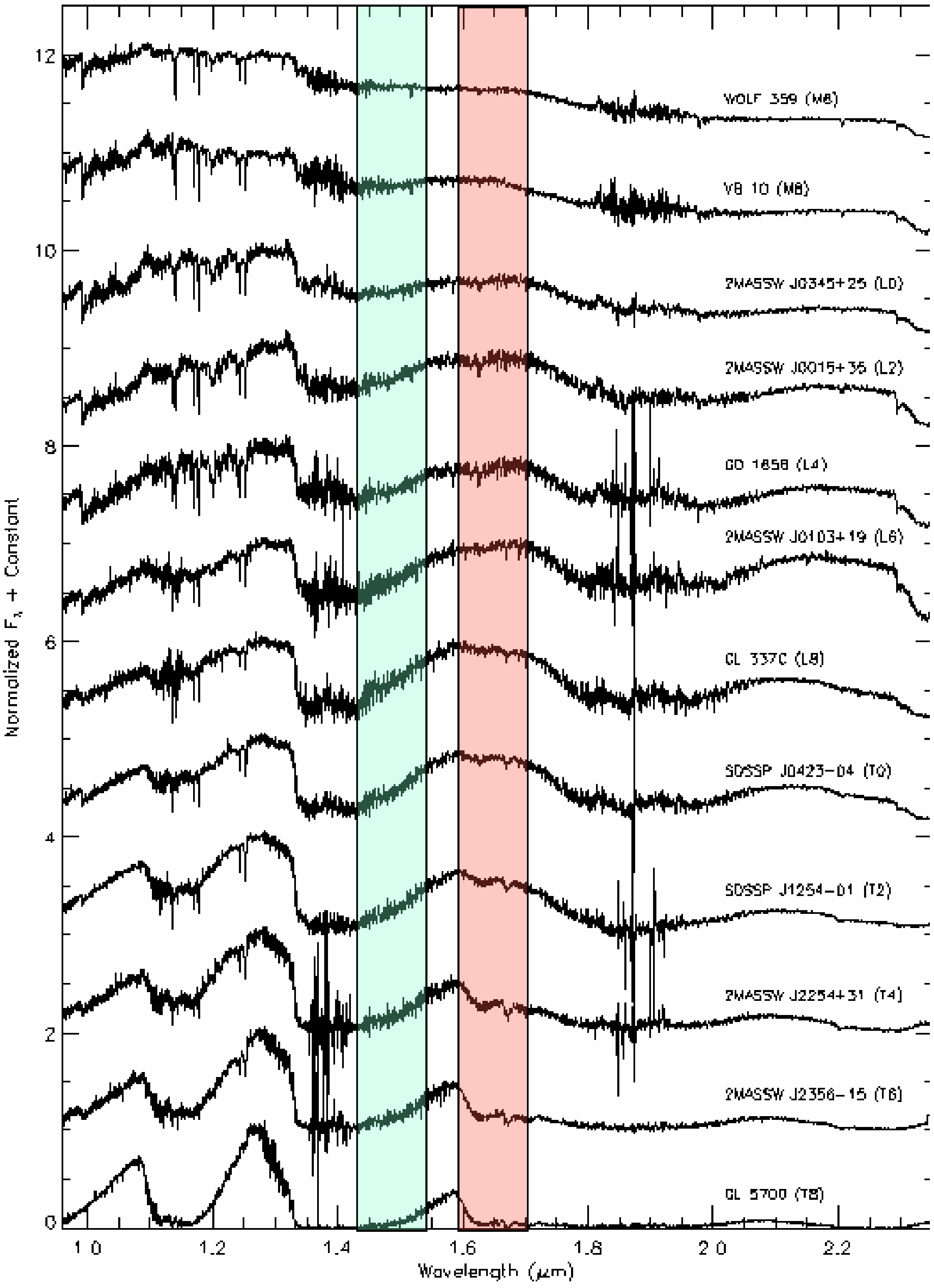

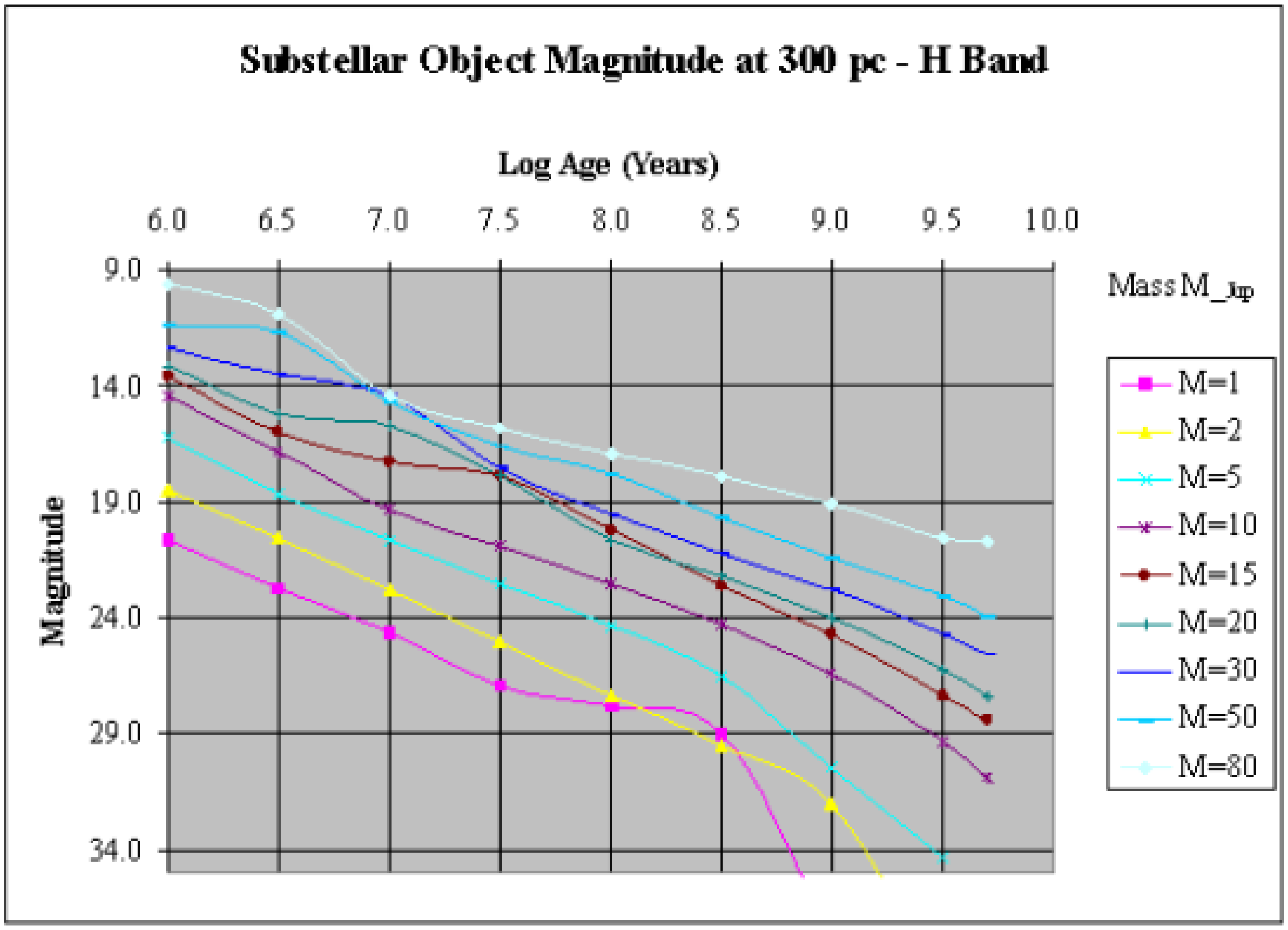

Substellar Object Magnitude at 300 pc - H Band
Log Age (Years)
6.0
6.5
7.0
7.5
8.0
8.5
9.0
9.5
10.0
9.0
14.0
19.0
24.0
29.0
34.0
Magnitude
Mass M_Jup
M=1
M=2
M=5
M=10
M=15
M=20
M=30
M=50
M=80

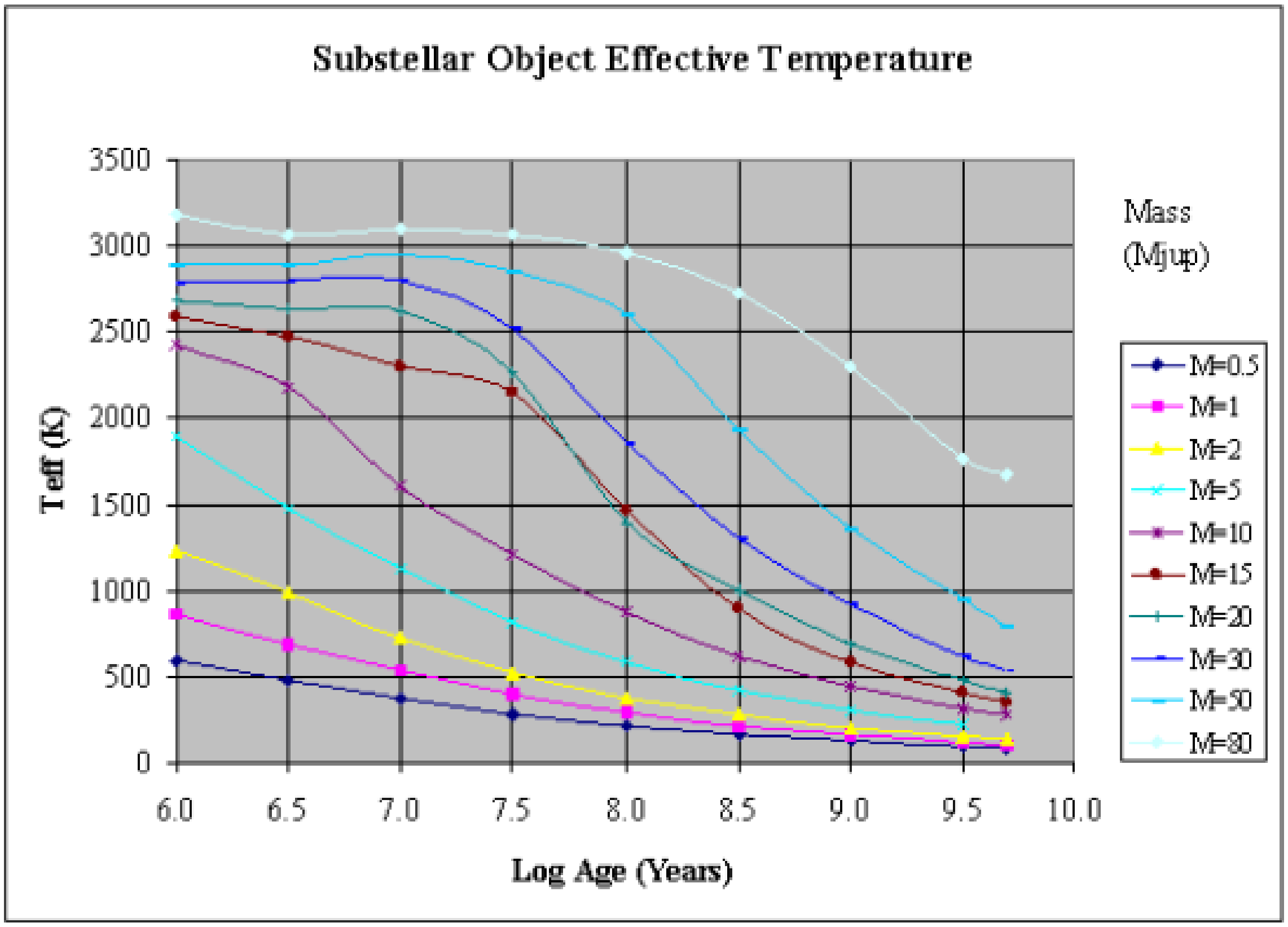

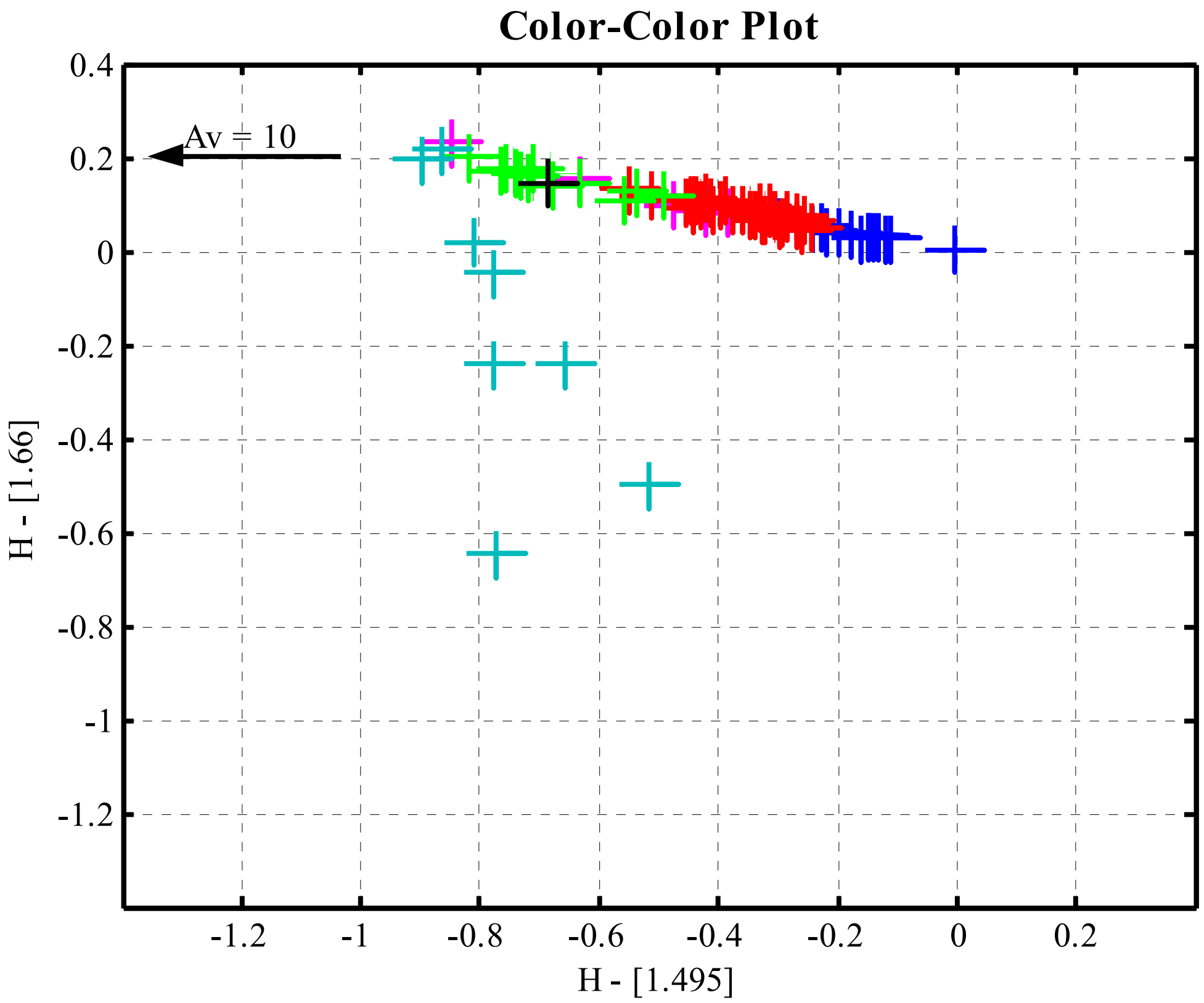

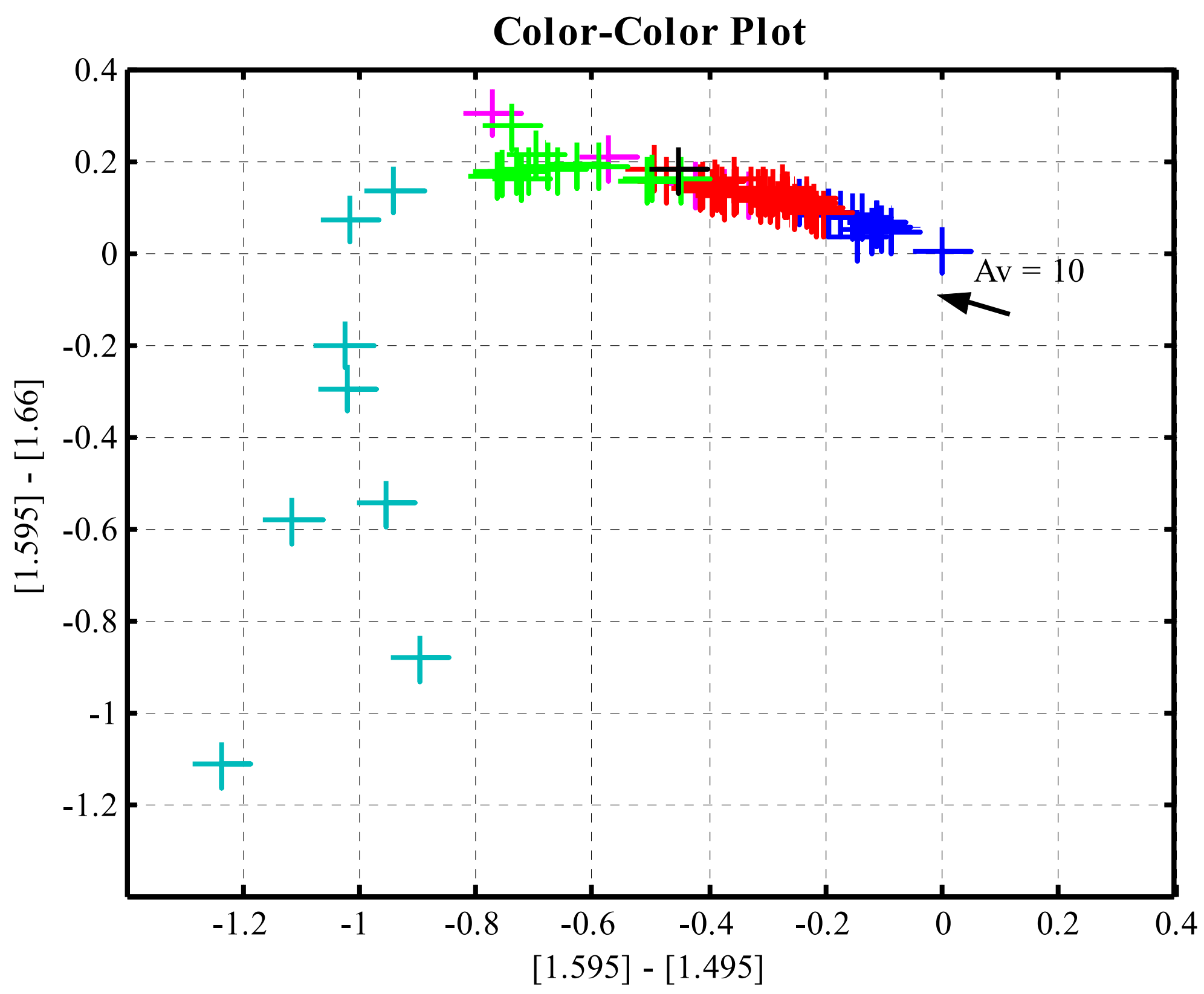

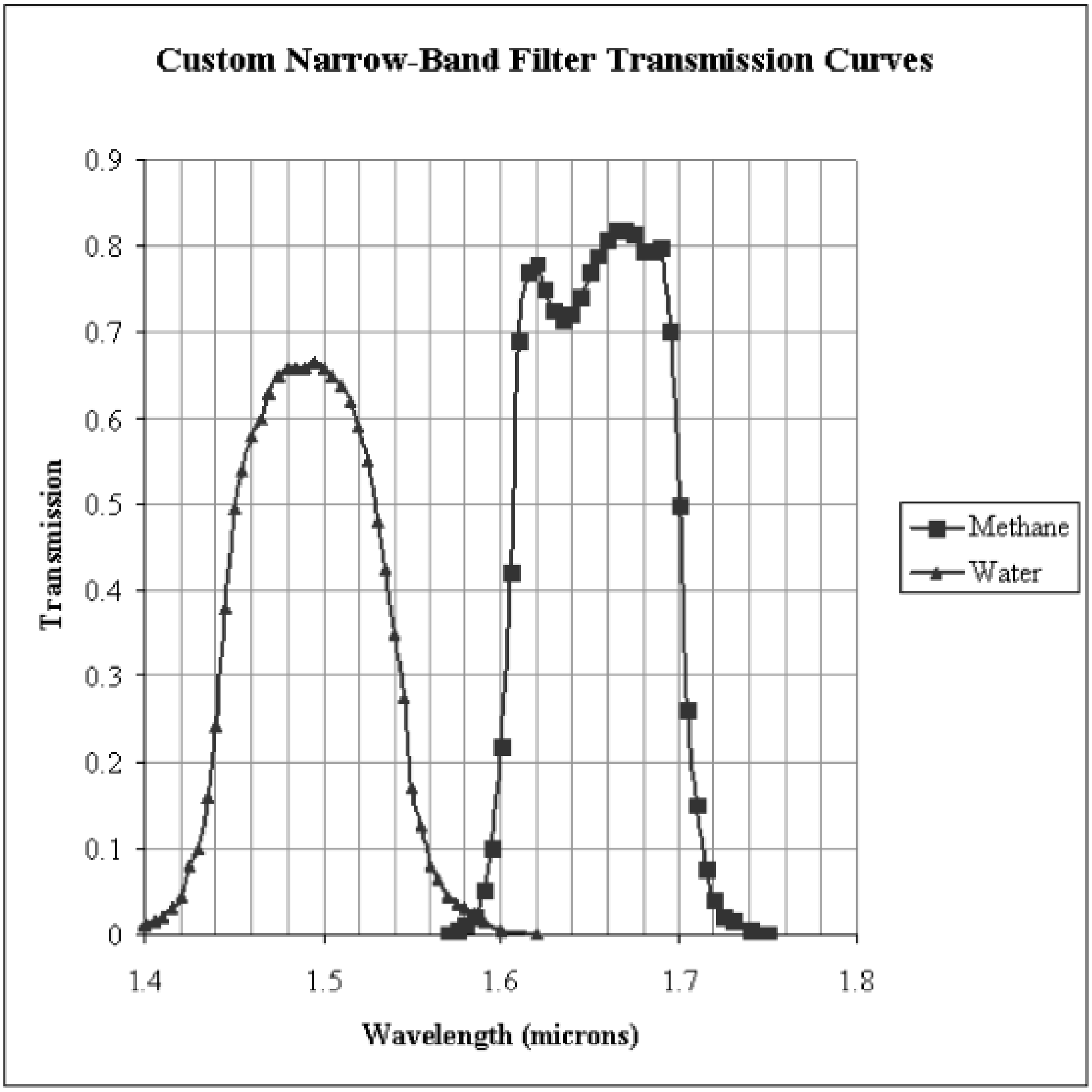

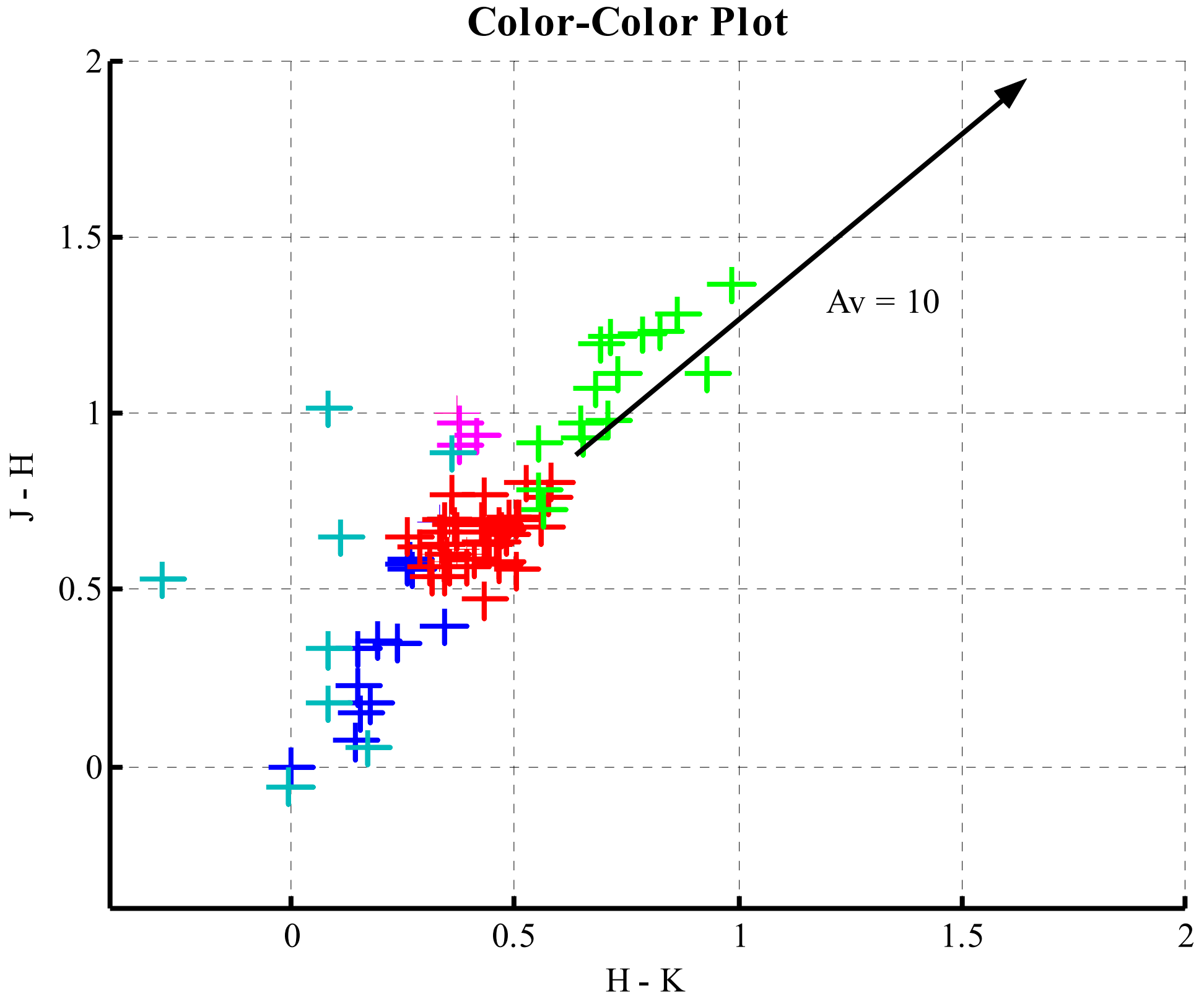

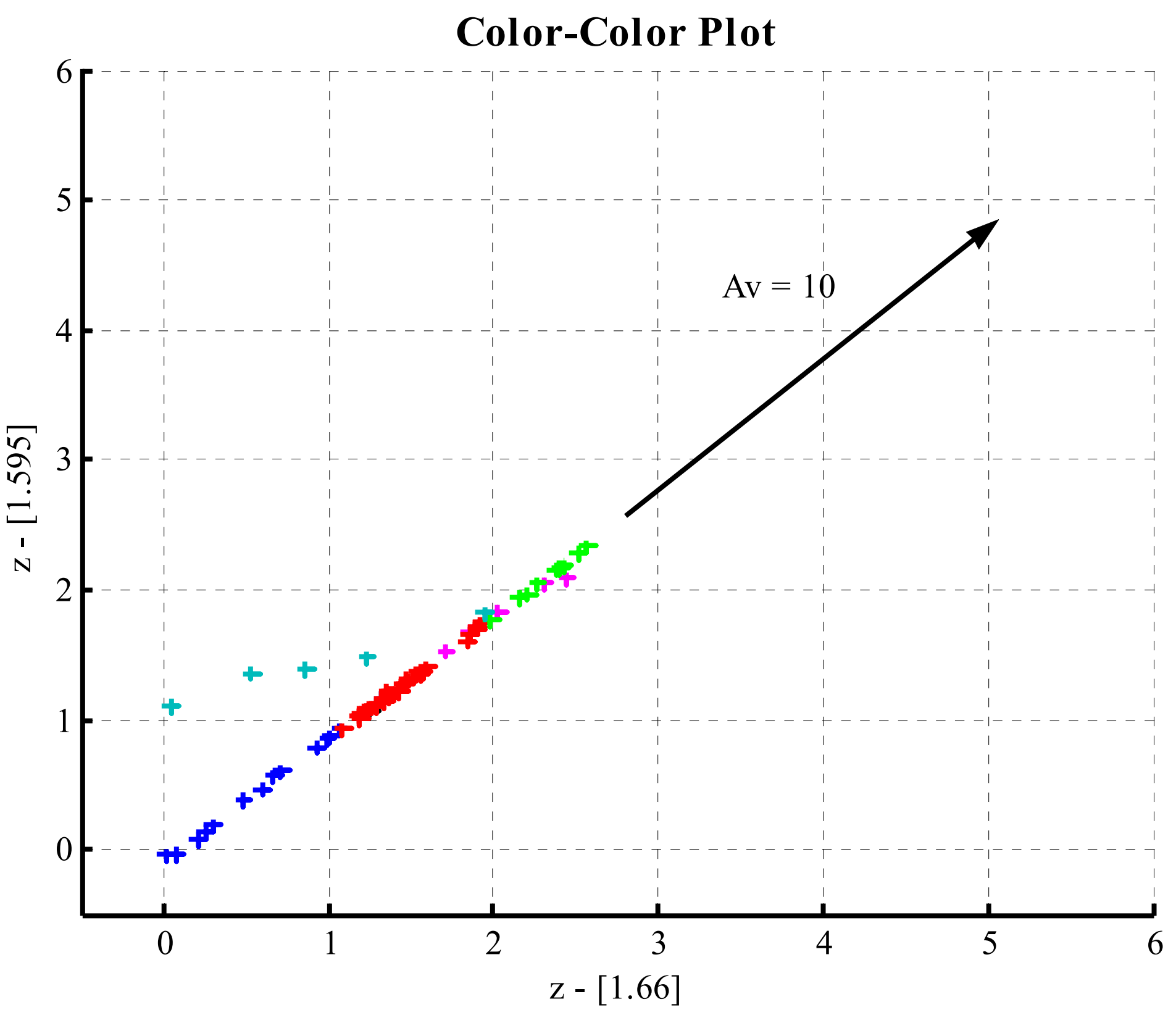

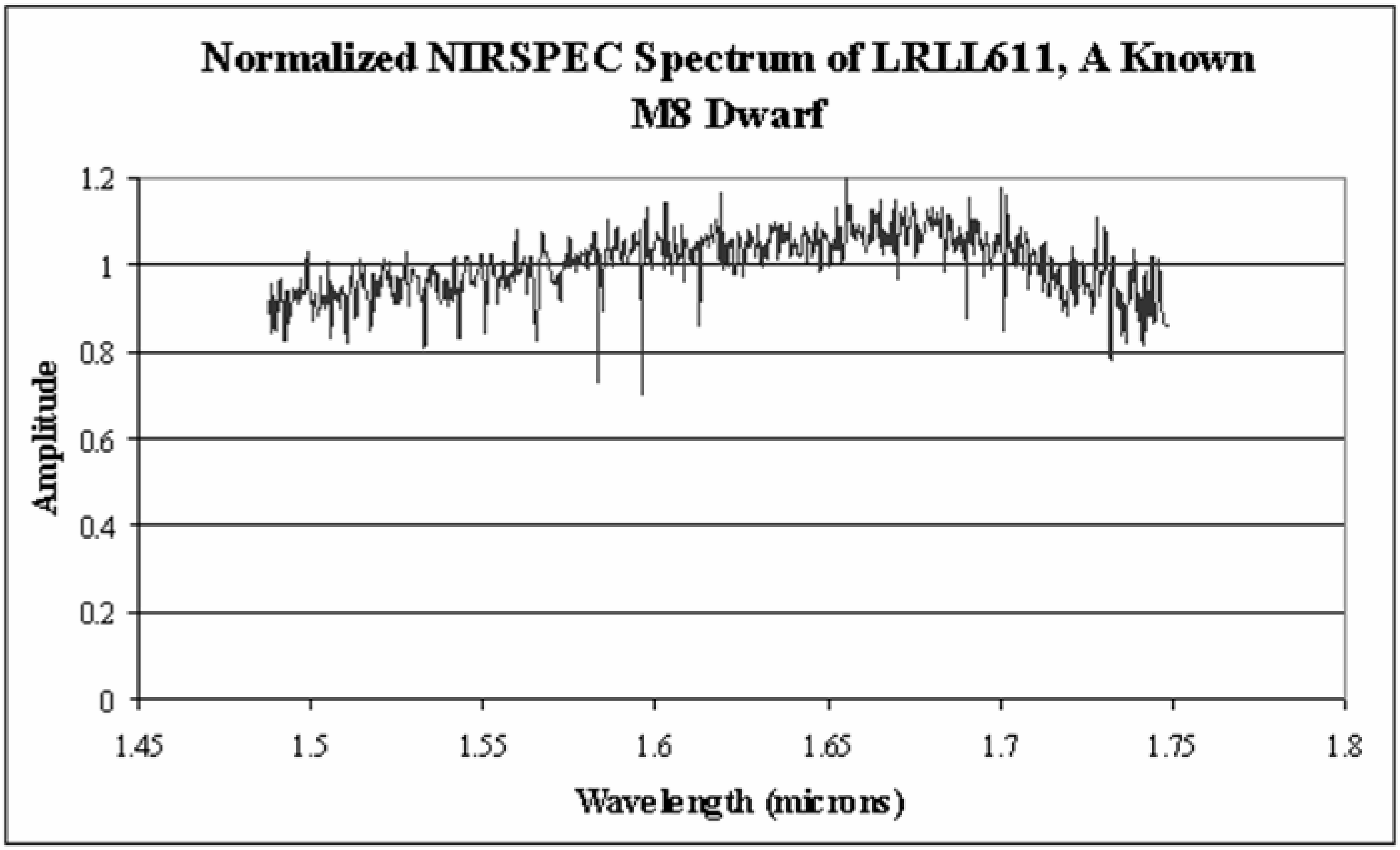
Normalized NIRSPEC Spectrum of LRLL611, A Known M8 Dwarf
Amplitude
1.2
1
0.8
0.6
0.4
0.2
0
1.45
1.5
1.55
1.6
1.65
1.7
1.75
1.8
Wavelength (microns)

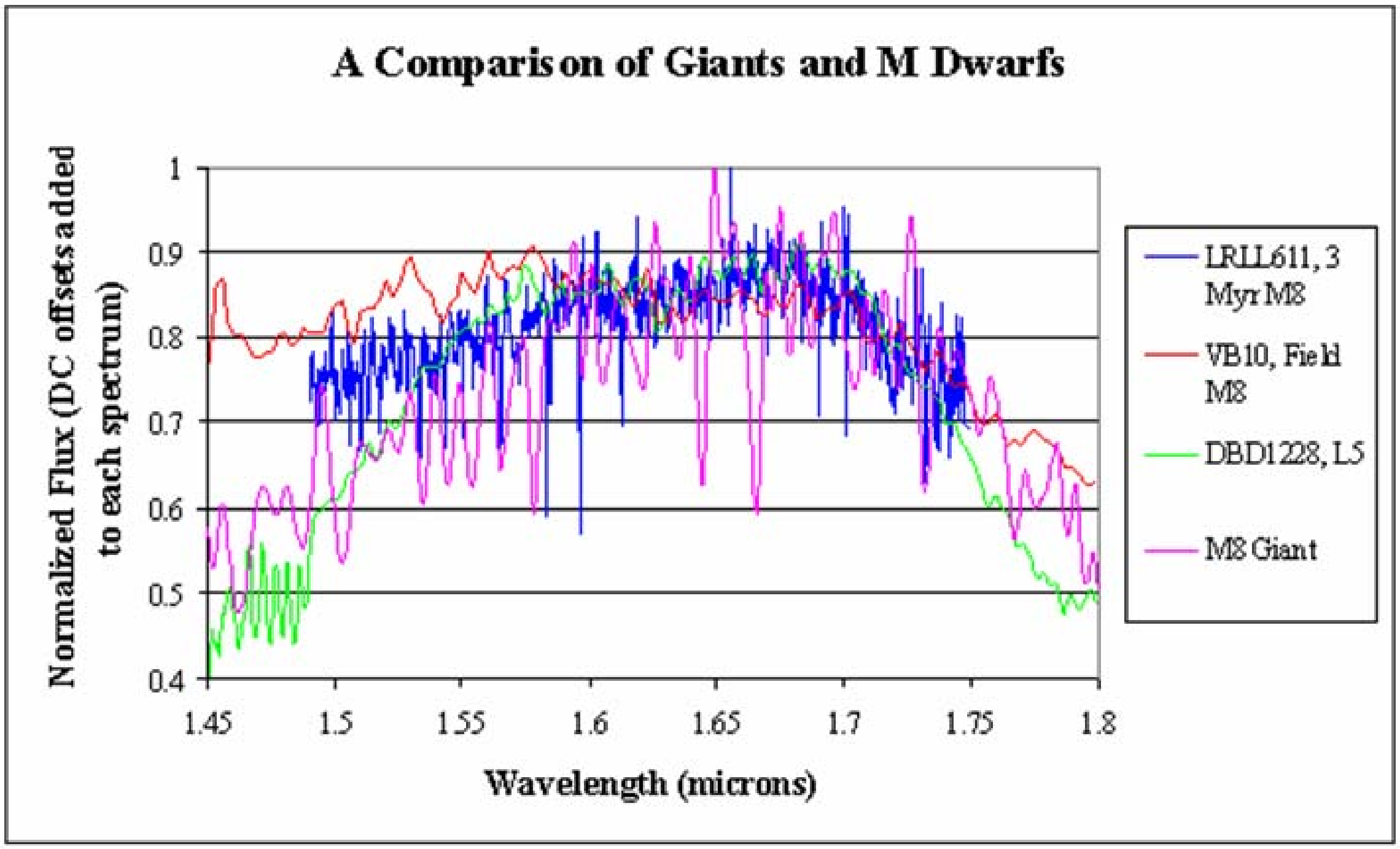
A Comparison of Giants and M Dwarfs
Normalized Flux (DC offsets added to each spectrum)
1
0.9
0.8
0.7
0.6
0.5
0.4
1.45
1.5
1.55
1.6
1.65
1.7
1.75
1.8
Wavelength (microns)
LRLL611, 3 Myr M8
VB10, Field M8
DBD1228, L5
M8 Giant

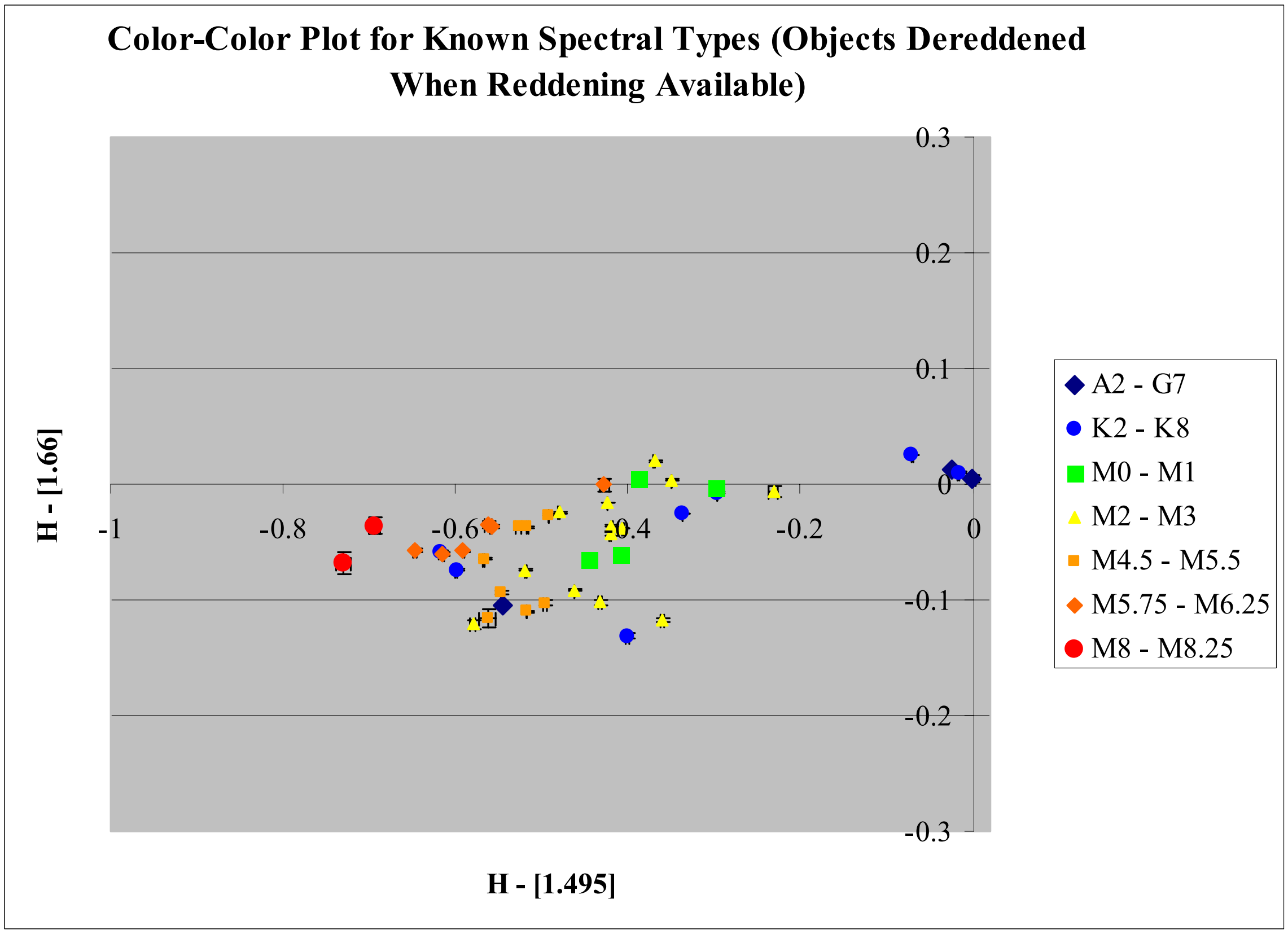
Color-Color Plot for Known Spectral Types (Objects Dereddened When Reddening Available)
0.3
0.2
0.1
0
-0.1
-0.2
-0.3
-1
-0.8
-0.6
-0.4
-0.2
0
H - [1.66]
H - [1.495]
A2 - G7
K2 - K8
M0 - M1
M2 - M3
M4.5 - M5.5
M5.75 - M6.25
M8 - M8.25

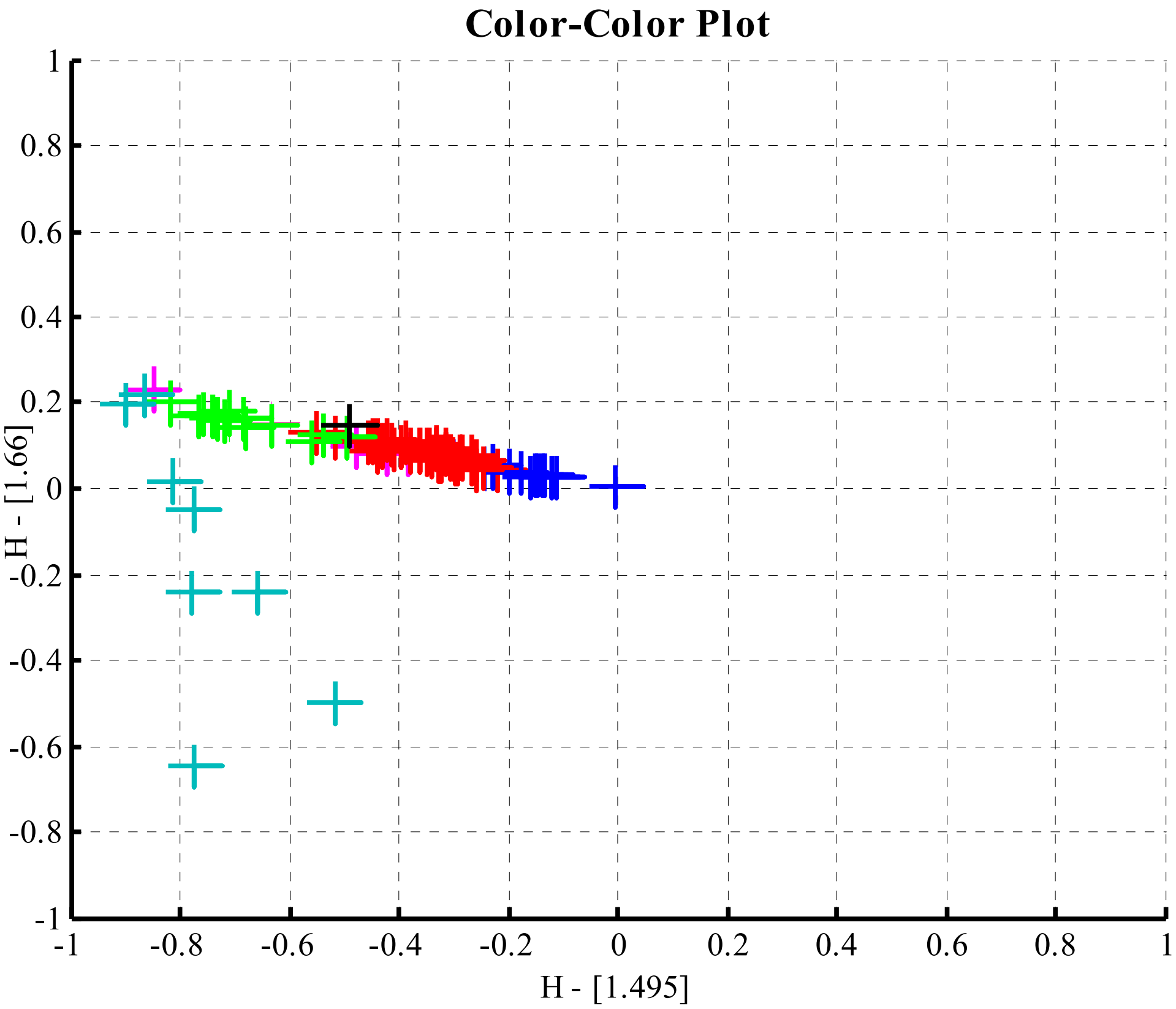

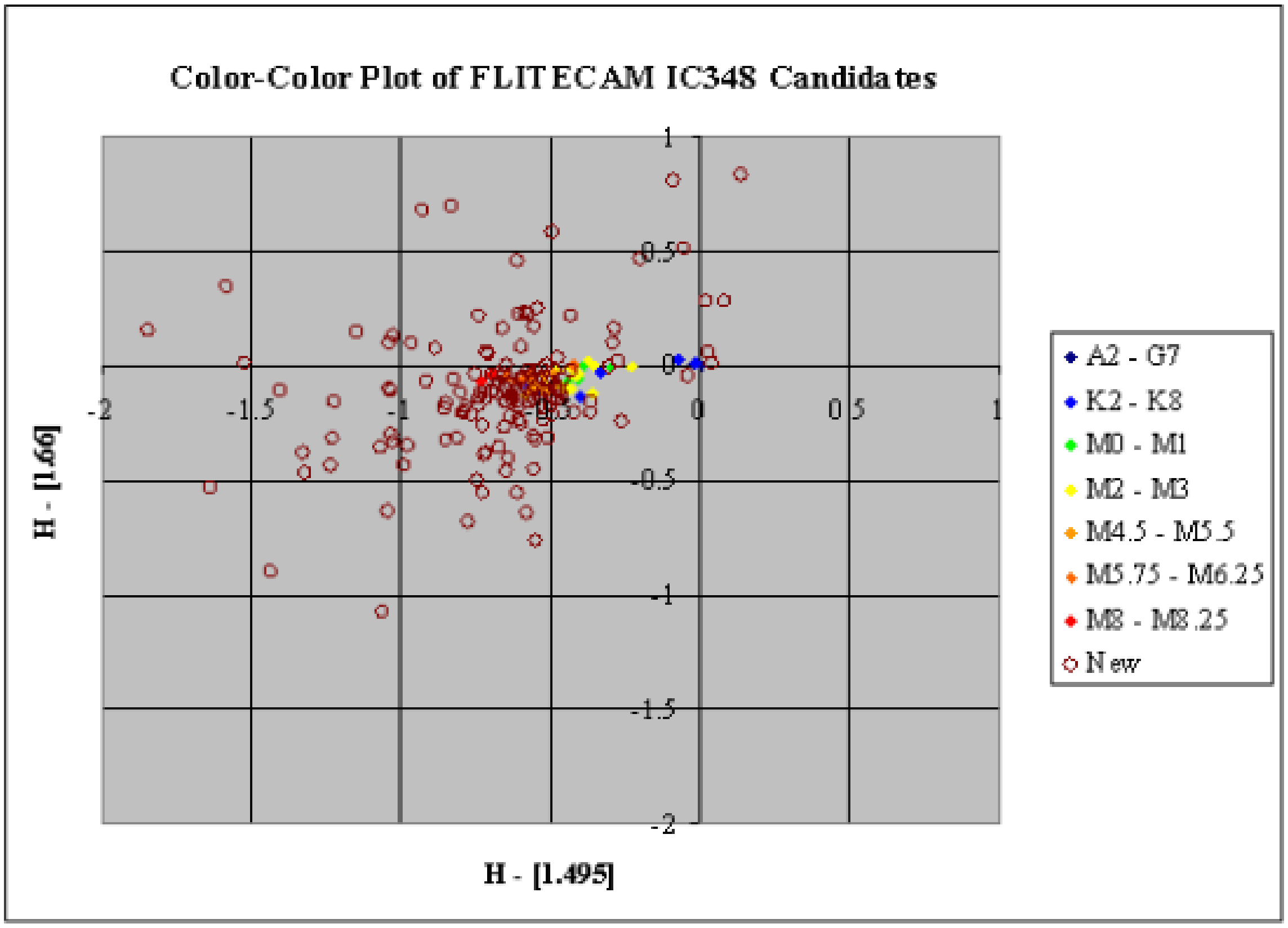

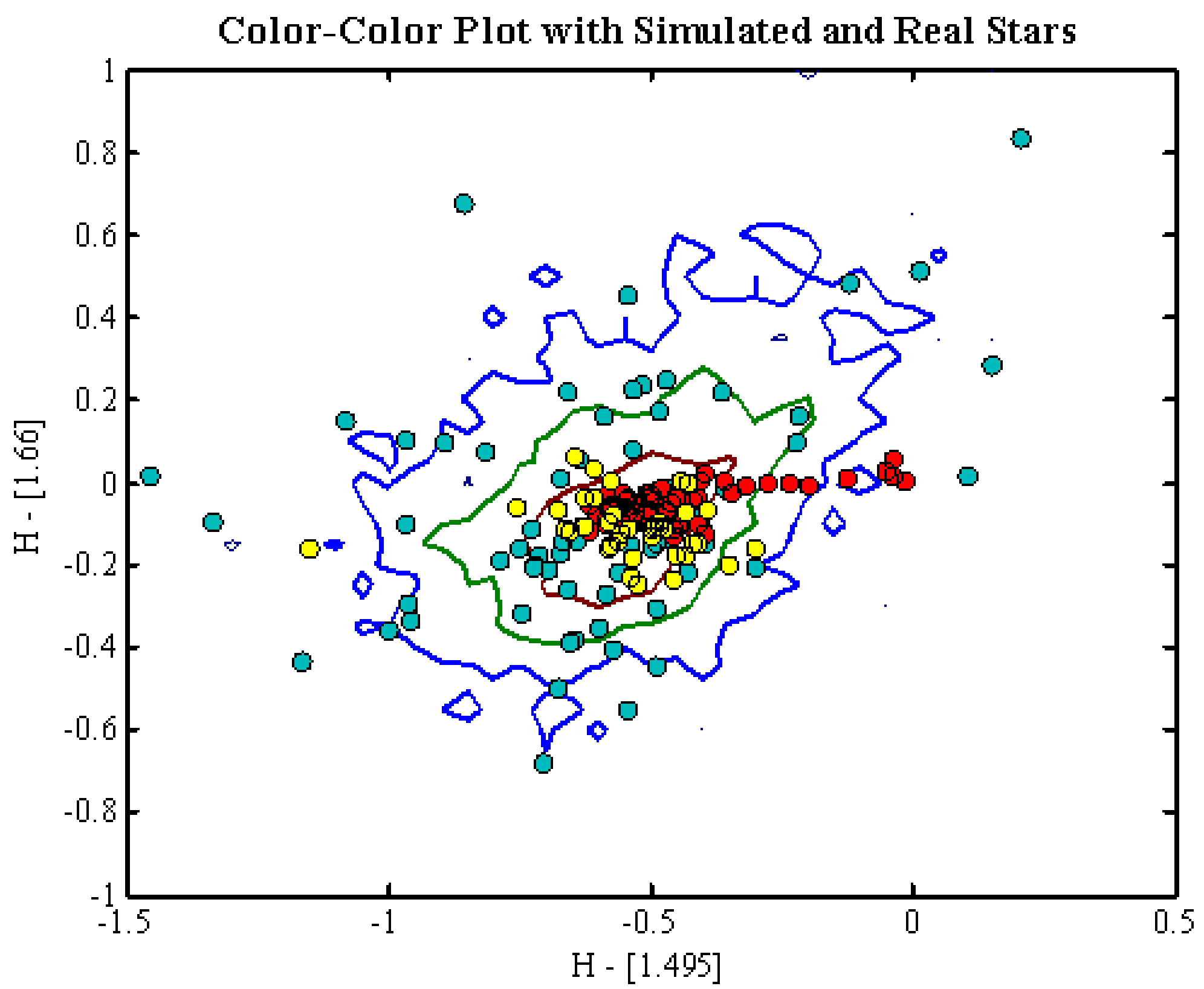

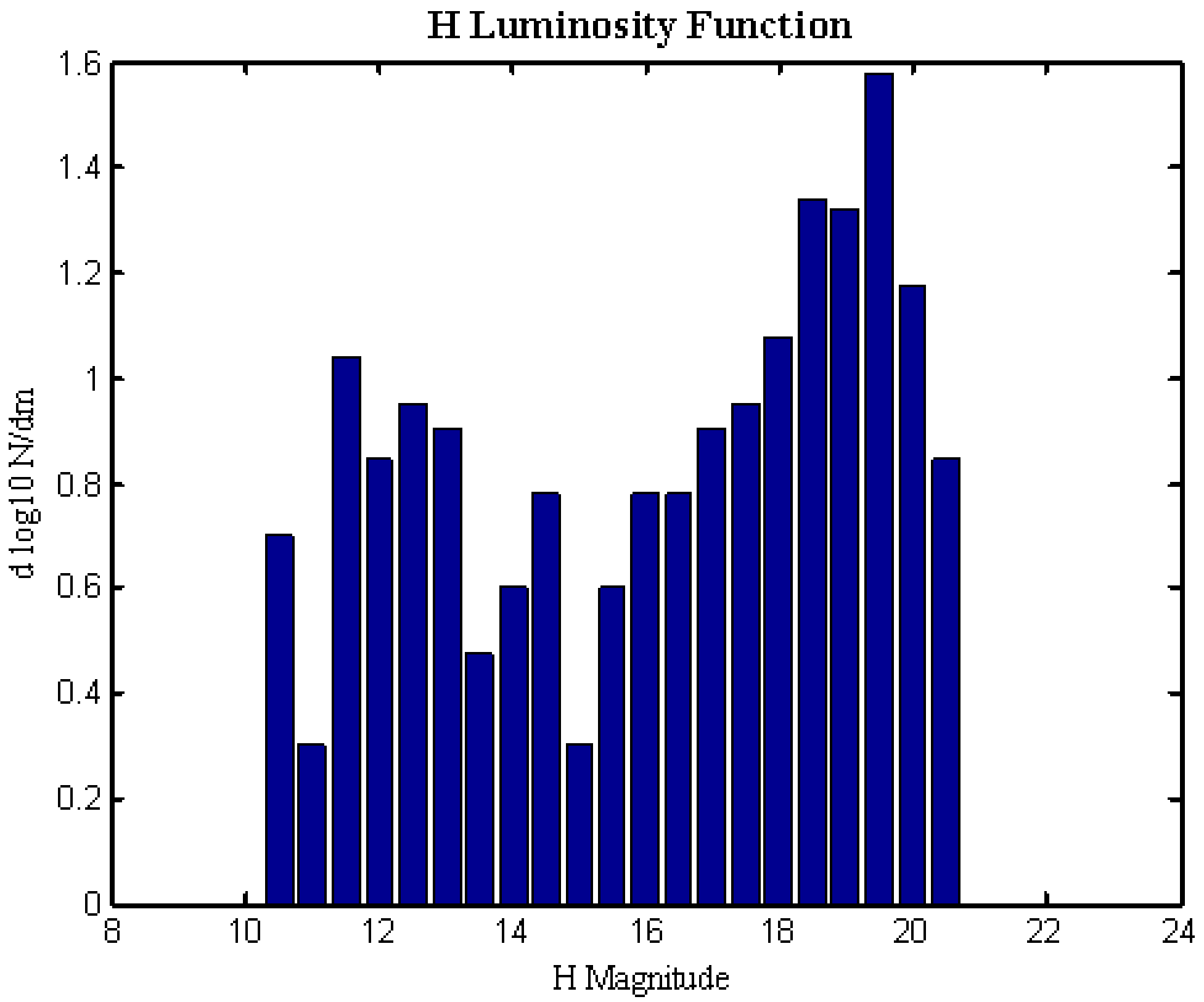

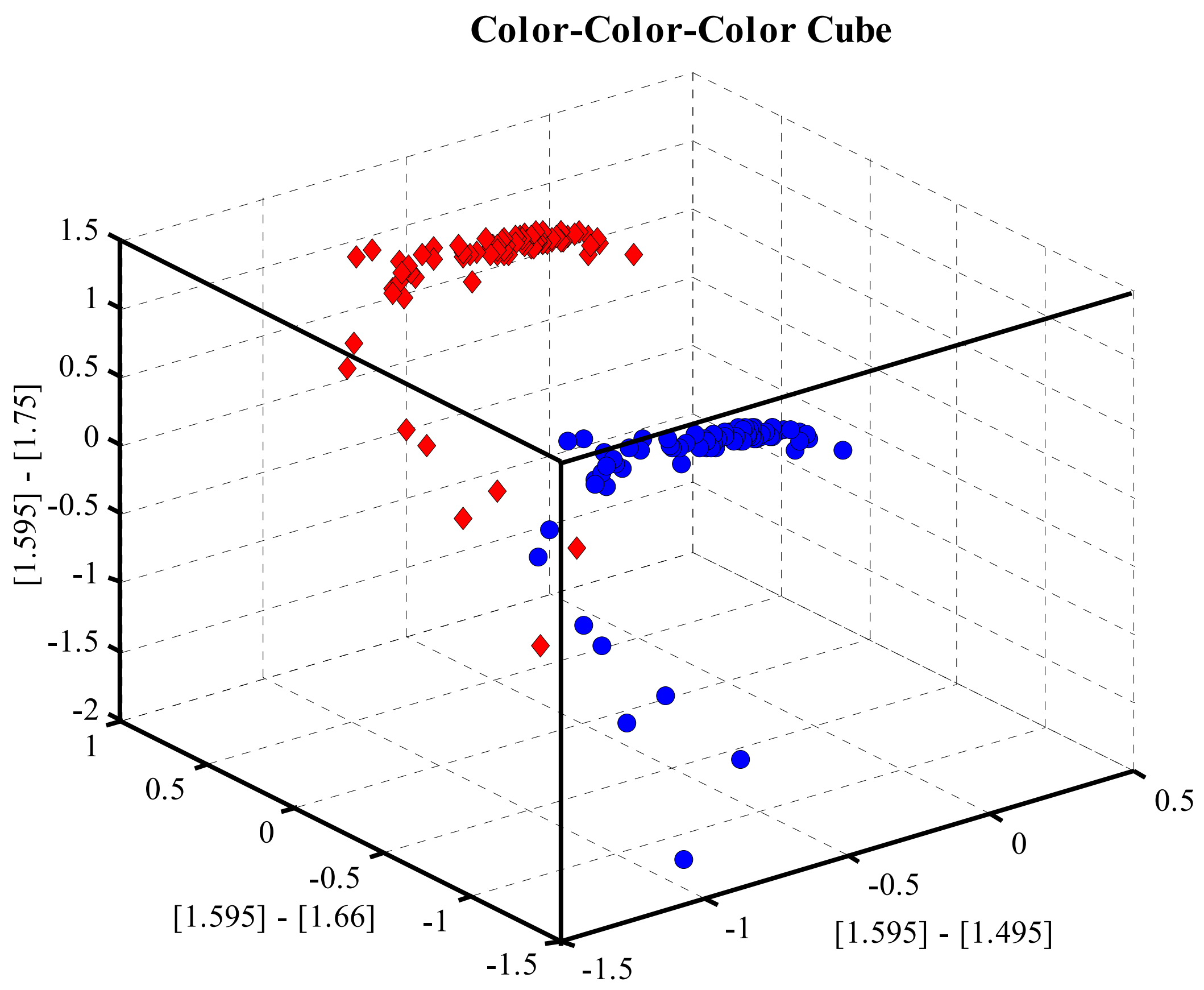

Color-Color-Color Cube, Rotated to Remove Reddening

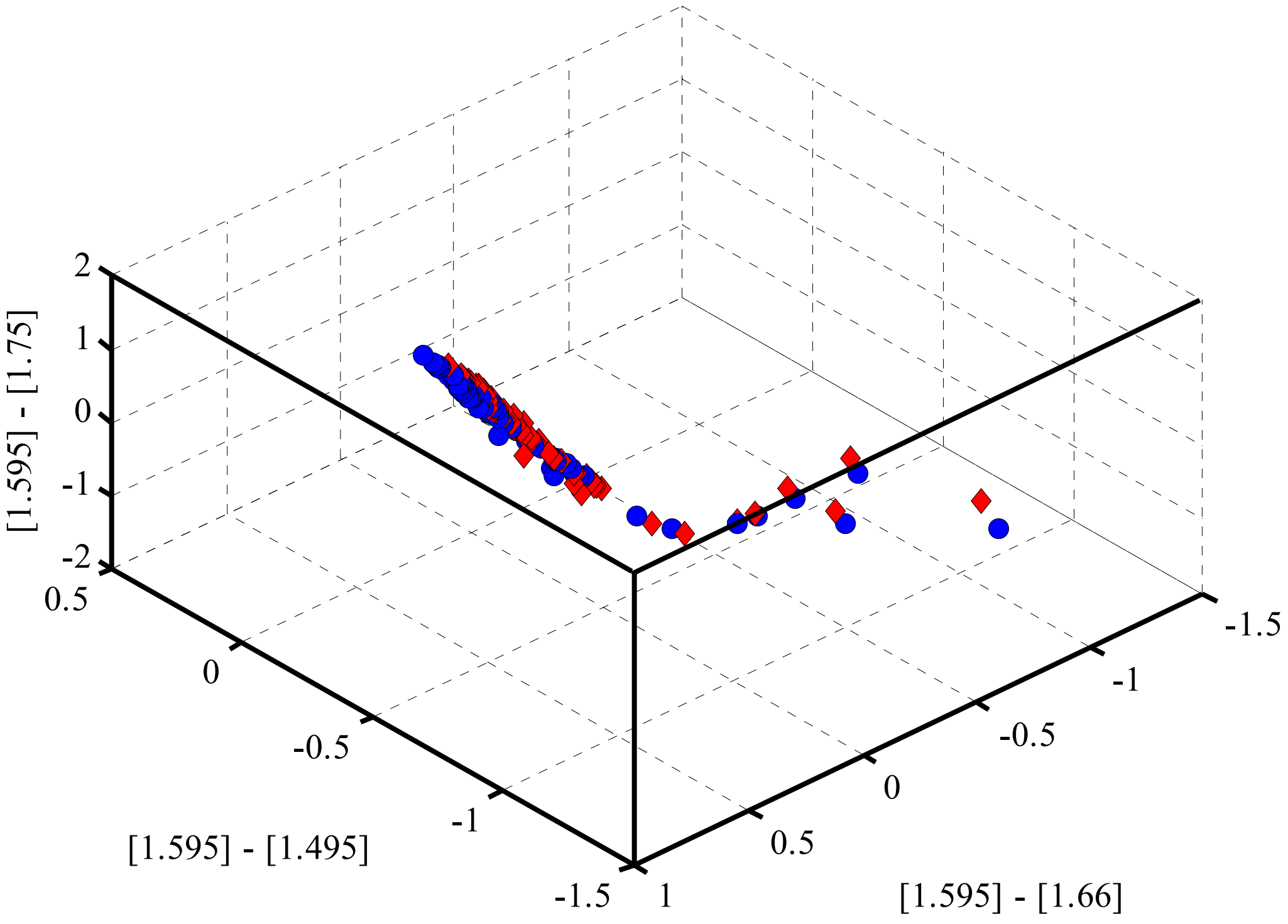

| Object* | Other ID* | Spectral Type | RA(J2000) | | | Dec (J2000) | | | H | H Err | [1.66] | [1.66] Err | [1.495] | [1.495] Err |
|---|---|---|---|---|---|---|---|---|---|---|---|---|---|---|
| H140 | | | 3 | 44 | 32.37 | 32 | 11 | 42.8 | 10.63 | 0.02 | 10.64 | 0.03 | 10.79 | 0.02 |
| H195 | | | 3 | 44 | 41.32 | 32 | 10 | 07.9 | 12.42 | 0.02 | 12.47 | 0.02 | 12.87 | 0.02 |
| H198 | ACIS-163 | | 3 | 44 | 41.45 | 32 | 10 | 23.4 | 11.88 | 0.02 | 11.93 | 0.02 | 12.26 | 0.02 |
| H204 | | | 3 | 44 | 42.69 | 32 | 10 | 00.3 | 12.46 | 0.02 | 12.49 | 0.02 | 12.96 | 0.02 |
| H207 | ACIS-167 | | 3 | 44 | 43.20 | 32 | 10 | 13.2 | 12.71 | 0.02 | 12.76 | 0.02 | 13.13 | 0.02 |
| H214 | ACIS-214 | | 3 | 44 | 43.93 | 32 | 10 | 28.7 | 11.49 | 0.02 | 11.61 | 0.02 | 11.87 | 0.02 |
| H218 | | | 3 | 44 | 44.61 | 32 | 10 | 03.8 | 14.48 | 0.02 | 14.58 | 0.02 | 14.81 | 0.02 |
| H257 | | | 3 | 44 | 43.43 | 32 | 09 | 38.3 | 16.20 | 0.02 | 16.32 | 0.02 | 16.67 | 0.02 |
| H258 | | | 3 | 44 | 43.78 | 32 | 09 | 04.4 | 17.58 | 0.02 | 17.73 | 0.02 | 17.92 | 0.02 |
| H259 | | | 3 | 44 | 46.68 | 32 | 08 | 59.6 | 14.48 | 0.02 | 14.56 | 0.02 | 15.00 | 0.02 |
| LRLL108 | ACIS-148 | M2 | 3 | 44 | 38.77 | 32 | 08 | 54.6 | 11.70 | 0.02 | 11.74 | 0.02 | 12.03 | 0.02 |
| LRLL113 | ACIS-135 | K5.5 | 3 | 44 | 37.31 | 32 | 09 | 13.8 | 11.76 | 0.02 | 11.78 | 0.02 | 12.02 | 0.02 |
| LRLL115 | ACIS-89 | M1 | 3 | 44 | 30.20 | 32 | 09 | 19.6 | 12.05 | 0.02 | 12.11 | 0.02 | 12.56 | 0.02 |
| LRLL128 | | M2 | 3 | 44 | 20.36 | 32 | 08 | 56.8 | 12.28 | 0.02 | 12.38 | 0.02 | 12.64 | 0.02 |
| LRLL151 | ACIS-118 | M2 | 3 | 44 | 35.00 | 32 | 11 | 17.4 | 12.20 | 0.02 | 12.25 | 0.02 | 12.54 | 0.02 |
| LRLL153 | | M2 | 3 | 44 | 42.80 | 32 | 08 | 31.7 | 12.23 | 0.02 | 12.25 | 0.02 | 12.65 | 0.02 |
| LRLL158 | ACIS-158 | M5 | 3 | 44 | 40.24 | 32 | 09 | 10.8 | 12.30 | 0.02 | 12.32 | 0.02 | 12.77 | 0.02 |
| LRLL16 | ACIS-108 | G6 | 3 | 44 | 32.92 | 32 | 08 | 35.9 | 10.63 | 0.02 | 10.63 | 0.02 | 10.56 | 0.02 |
| LRLL165 | | M5 | 3 | 44 | 35.65 | 32 | 08 | 54.1 | 12.40 | 0.02 | 12.44 | 0.02 | 12.87 | 0.02 |
| LRLL168 | | M2 | 3 | 44 | 31.55 | 32 | 10 | 46.0 | 12.53 | 0.02 | 12.55 | 0.02 | 12.93 | 0.02 |
| LRLL170 | ACIS-82 | M3 | 3 | 44 | 28.66 | 32 | 11 | 22.5 | 12.93 | 0.02 | 13.05 | 0.02 | 13.47 | 0.02 |
| LRLL19 | ACIS-94 | A2 | 3 | 44 | 31.03 | 32 | 09 | 54.9 | 10.64 | 0.02 | 10.63 | 0.02 | 10.59 | 0.02 |
| LRLL191 | ACIS-141 | M2 | 3 | 44 | 38.00 | 32 | 10 | 05.3 | 12.58 | 0.02 | 12.67 | 0.02 | 13.06 | 0.02 |
| LRLL194 | | M1 | 3 | 44 | 27.44 | 32 | 10 | 37.2 | 12.83 | 0.02 | 12.90 | 0.02 | 13.22 | 0.02 |
| LRLL221 | | K5 | 3 | 44 | 40.35 | 32 | 09 | 30.9 | 12.92 | 0.02 | 12.99 | 0.02 | 13.38 | 0.02 |
| LRLL226 | ACIS-101 | M3 | 3 | 44 | 31.66 | 32 | 11 | 28.8 | 13.00 | 0.02 | 13.08 | 0.02 | 13.45 | 0.02 |
| LRLL23 | ACIS-149 | K3 | 3 | 44 | 38.81 | 32 | 08 | 40.2 | 10.66 | 0.02 | 10.65 | 0.02 | 10.70 | 0.02 |
| LRLL232 | | F7 – G7 | 3 | 44 | 36.59 | 32 | 09 | 17.5 | 13.11 | 0.02 | 13.21 | 0.02 | 13.50 | 0.02 |
| LRLL237 | | M5 | 3 | 44 | 23.74 | 32 | 09 | 33.7 | 12.93 | 0.02 | 13.04 | 0.02 | 13.35 | 0.02 |
| LRLL248 | ACIS-130 | K5 | 3 | 44 | 36.09 | 32 | 09 | 22.2 | 13.07 | 0.02 | 13.13 | 0.02 | 13.54 | 0.02 |
| LRLL277 | ACIS-154 | M5 | 3 | 44 | 39.59 | 32 | 10 | 05.9 | 13.08 | 0.02 | 13.14 | 0.02 | 13.50 | 0.02 |
| LRLL29 | ACIS-103 | K2 | 3 | 44 | 31.72 | 32 | 08 | 43.5 | 10.78 | 0.02 | 10.75 | 0.02 | 10.75 | 0.02 |
| LRLL294 | | M4.5 | 3 | 44 | 24.76 | 32 | 10 | 02.8 | 13.63 | 0.02 | 13.73 | 0.02 | 13.99 | 0.02 |
| LRLL325 | | M6 | 3 | 44 | 30.28 | 32 | 08 | 47.6 | 13.76 | 0.02 | 13.81 | 0.02 | 14.25 | 0.02 |
| LRLL33 | ACIS-106 | M3 | 3 | 44 | 32.76 | 32 | 08 | 41.0 | 10.94 | 0.02 | 10.95 | 0.02 | 11.06 | 0.02 |
| LRLL335 | | M5.75 | 3 | 44 | 44.30 | 32 | 08 | 45.0 | 13.76 | 0.02 | 13.82 | 0.02 | 14.26 | 0.02 |
| LRLL351 | | M3 | 3 | 44 | 25.96 | 32 | 09 | 05.1 | 13.75 | 0.02 | 13.85 | 0.02 | 14.26 | 0.02 |
| LRLL353 | | M6 | 3 | 44 | 38.32 | 32 | 10 | 19.6 | 13.67 | 0.02 | 13.71 | 0.02 | 14.09 | 0.02 |
| LRLL40 | ACIS-88 | K8 | 3 | 44 | 29.91 | 32 | 10 | 39.6 | 11.29 | 0.02 | 11.30 | 0.02 | 11.53 | 0.02 |
| LRLL41 | ACIS-51 | K7 | 3 | 44 | 21.86 | 32 | 10 | 38.3 | 12.17 | 0.02 | 12.30 | 0.02 | 12.55 | 0.02 |
| LRLL414 | | M5.25 | 3 | 44 | 44.46 | 32 | 10 | 35.1 | 14.90 | 0.02 | 15.02 | 0.02 | 15.32 | 0.02 |
| LRLL415 | | M6.25 | 3 | 44 | 30.18 | 32 | 09 | 38.0 | 14.28 | 0.02 | 14.34 | 0.02 | 14.82 | 0.02 |
| LRLL435 | ACIS-91 | | 3 | 44 | 30.56 | 32 | 11 | 34.6 | 15.47 | 0.02 | 15.60 | 0.02 | 15.95 | 0.02 |
| LRLL442 | | | 3 | 44 | 40.76 | 32 | 09 | 38.8 | 14.45 | 0.02 | 14.58 | 0.02 | 14.77 | 0.02 |
| LRLL454 | | M5.75 | 3 | 44 | 41.79 | 32 | 10 | 37.8 | 14.74 | 0.02 | 14.78 | 0.02 | 15.16 | 0.02 |
| LRLL478 | | M6.25 | 3 | 44 | 36.18 | 32 | 11 | 16.6 | 15.29 | 0.02 | 15.29 | 0.02 | 15.63 | 0.02 |
| LRLL609 | | | 3 | 44 | 45.01 | 32 | 09 | 34.9 | 17.33 | 0.02 | 17.46 | 0.02 | 17.75 | 0.02 |
| LRLL611 | | M8 | 3 | 44 | 30.57 | 32 | 09 | 43.0 | 15.59 | 0.02 | 15.62 | 0.02 | 16.13 | 0.02 |
| LRLL613 | | M8..25 | 3 | 44 | 27.10 | 32 | 09 | 25.2 | 16.01 | 0.02 | 16.08 | 0.02 | 16.61 | 0.02 |
| LRLL618 | | | 3 | 44 | 43.98 | 32 | 08 | 34.5 | 16.84 | 0.02 | 16.99 | 0.02 | 17.32 | 0.02 |
| LRLL65 | | M0 | 3 | 44 | 34.18 | 32 | 08 | 52.4 | 11.32 | 0.02 | 11.32 | 0.02 | 11.51 | 0.02 |
| LRLL71 | ACIS-107 | M3 | 3 | 44 | 32.78 | 32 | 08 | 53.9 | 11.41 | 0.02 | 11.41 | 0.02 | 11.69 | 0.02 |
| LRLL74 | ACIS-115 | M2 | 3 | 44 | 34.49 | 32 | 10 | 48.7 | 11.42 | 0.02 | 11.40 | 0.02 | 11.73 | 0.02 |
| LRLL78 | | M3 | 3 | 44 | 26.88 | 32 | 08 | 20.4 | 11.75 | 0.02 | 11.82 | 0.02 | 12.21 | 0.02 |
| LRLL83 | ACIS-138 | M1 | 3 | 44 | 37.59 | 32 | 08 | 58.7 | 11.65 | 0.02 | 11.64 | 0.02 | 11.97 | 0.02 |
| LRLL91 | ACIS-152 | M2 | 3 | 44 | 39.32 | 32 | 09 | 42.9 | 11.54 | 0.02 | 11.57 | 0.02 | 11.91 | 0.02 |
| MM1 | NTC071-04 | | 3 | 44 | 38.55 | 32 | 10 | 46.3 | 17.56 | 0.03 | 17.63 | 0.03 | 17.87 | 0.03 |
| MM10 | NTC071-05 | | 3 | 44 | 38.33 | 32 | 10 | 30.1 | 18.04 | 0.04 | 18.13 | 0.04 | 18.37 | 0.04 |
| MM11 | | | 3 | 44 | 35.62 | 32 | 11 | 03.4 | 19.18 | 0.11 | 18.94 | 0.11 | 19.62 | 0.11 |
| MM12 | NTC084-07 | | 3 | 44 | 34.25 | 32 | 11 | 02.0 | 18.69 | 0.06 | 18.71 | 0.07 | 19.03 | 0.06 |
| MM13 | NTC084-04 | | 3 | 44 | 33.73 | 32 | 11 | 18.2 | 17.86 | 0.03 | 18.07 | 0.04 | 18.48 | 0.04 |
| MM14 | | | 3 | 44 | 30.14 | 32 | 09 | 48.1 | 20.12 | 0.19 | 19.93 | 0.16 | 22.60 | 1.06 |
| MM15 | NTC072-02 | | 3 | 44 | 38.77 | 32 | 11 | 02.8 | 16.91 | 0.02 | 17.07 | 0.02 | 17.41 | 0.02 |
| MM16 | NTC064-02 | | 3 | 44 | 43.09 | 32 | 10 | 28.0 | 17.00 | 0.02 | 17.15 | 0.02 | 17.49 | 0.02 |
| MM17 | NTC064-07 | | 3 | 44 | 44.32 | 32 | 09 | 41.8 | 17.50 | 0.02 | 17.68 | 0.03 | 17.88 | 0.02 |
| MM18 | | | 3 | 44 | 45.28 | 32 | 09 | 31.7 | 18.43 | 0.05 | 18.54 | 0.05 | 18.89 | 0.06 |
| MM19 | | | 3 | 44 | 46.79 | 32 | 09 | 26.2 | 19.25 | 0.08 | 19.46 | 0.12 | 19.47 | 0.09 |
| MM2 | NTC095-03 | | 3 | 44 | 27.43 | 32 | 10 | 06.2 | 17.12 | 0.02 | 17.22 | 0.02 | 17.66 | 0.02 |
| MM20 | | | 3 | 44 | 41.38 | 32 | 09 | 34.8 | 18.95 | 0.07 | 19.10 | 0.08 | 19.26 | 0.07 |
| MM21 | NTC024-02 | | 3 | 44 | 33.46 | 32 | 08 | 56.0 | 18.38 | 0.05 | 18.31 | 0.05 | 19.11 | 0.08 |
| MM22 | | | 3 | 44 | 41.73 | 32 | 09 | 10.7 | 19.76 | 0.15 | 19.54 | 0.13 | 20.19 | 0.19 |
| MM23 | | | 3 | 44 | 32.58 | 32 | 10 | 23.9 | 19.52 | 0.11 | 19.30 | 0.10 | 19.80 | 0.10 |
| MM24 | | | 3 | 44 | 35.46 | 32 | 10 | 22.7 | 19.92 | 0.21 | 19.44 | 0.13 | 19.96 | 0.17 |
| MM25 | | | 3 | 44 | 33.20 | 32 | 10 | 15.3 | 19.05 | 0.09 | 19.49 | 0.12 | 20.14 | 0.19 |
| MM26 | | | 3 | 44 | 31.72 | 32 | 10 | 35.5 | 19.10 | 0.09 | 19.43 | 0.10 | 19.98 | 0.14 |
| MM27 | | | 3 | 44 | 35.41 | 32 | 11 | 04.1 | 19.23 | 0.10 | 19.07 | 0.09 | 19.74 | 0.14 |
| MM28 | | | 3 | 44 | 44.76 | 32 | 09 | 15.2 | 19.41 | 0.10 | 19.24 | 0.08 | 19.54 | 0.10 |
| MM29 | | | 3 | 44 | 48.53 | 32 | 08 | 56.9 | 18.23 | 0.04 | 18.54 | 0.06 | 18.64 | 0.04 |
| MM3 | NTC102-02 | | 3 | 44 | 27.83 | 32 | 09 | 43.9 | 17.87 | 0.02 | 18.02 | 0.03 | 18.32 | 0.03 |
| MM30 | | | 3 | 44 | 36.73 | 32 | 09 | 01.6 | 18.55 | 0.05 | 18.82 | 0.07 | 19.05 | 0.06 |
| MM31 | | | 3 | 44 | 26.93 | 32 | 09 | 29.6 | 18.82 | 0.07 | 19.23 | 0.10 | 19.31 | 0.09 |
| MM32 | | | 3 | 44 | 40.44 | 32 | 10 | 38.2 | 18.38 | 0.05 | 18.52 | 0.06 | 18.77 | 0.07 |
| MM33 | | | 3 | 44 | 26.66 | 32 | 09 | 21.7 | 19.62 | 0.10 | 19.60 | 0.13 | 19.74 | 0.10 |
| MM34 | | | 3 | 44 | 24.16 | 32 | 09 | 07.8 | 18.91 | 0.06 | 19.13 | 0.07 | 19.39 | 0.08 |
| MM35 | | | 3 | 44 | 23.08 | 32 | 09 | 09.3 | 19.42 | 0.08 | 19.44 | 0.08 | 19.71 | 0.10 |
| MM36 | | | 3 | 44 | 27.95 | 32 | 08 | 49.9 | 18.83 | 0.06 | 19.02 | 0.06 | 19.53 | 0.08 |
| MM37 | | | 3 | 44 | 25.06 | 32 | 08 | 40.3 | 19.37 | 0.11 | 19.48 | 0.10 | 19.92 | 0.12 |
| MM38 | | | 3 | 44 | 26.84 | 32 | 08 | 33.4 | 18.50 | 0.05 | 18.72 | 0.07 | 18.85 | 0.06 |
| MM39 | | | 3 | 44 | 28.16 | 32 | 08 | 39.4 | 17.59 | 0.02 | 17.69 | 0.02 | 18.09 | 0.03 |

| | | | | | | | | | | | | | | | |
|---|---|---|---|---|---|---|---|---|---|---|---|---|---|---|---|
| MM4 | NTC013-03 | | 3 | 44 | 31.24 | 32 | 09 | 45.1 | 18.04 | 0.06 | 18.15 | 0.06 | 18.59 | 0.08 |
| MM40 | | | 3 | 44 | 27.78 | 32 | 08 | 42.7 | 16.36 | 0.02 | 16.48 | 0.02 | 16.93 | 0.02 |
| MM41 | | | 3 | 44 | 22.29 | 32 | 09 | 23.7 | 18.96 | 0.06 | 19.25 | 0.09 | 19.84 | 0.12 |
| MM42 | | | 3 | 44 | 21.88 | 32 | 09 | 18.7 | 19.20 | 0.08 | 19.35 | 0.10 | 19.61 | 0.08 |
| MM43 | | | 3 | 44 | 24.40 | 32 | 09 | 19.6 | 19.44 | 0.09 | 20.05 | 0.17 | 20.33 | 0.14 |
| MM44 | | | 3 | 44 | 23.92 | 32 | 09 | 21.1 | 20.00 | 0.15 | 20.24 | 0.19 | 20.12 | 0.16 |
| MM45 | | | 3 | 44 | 41.79 | 32 | 08 | 32.5 | 18.80 | 0.06 | 18.96 | 0.07 | 19.22 | 0.08 |
| MM46 | | | 3 | 44 | 40.59 | 32 | 08 | 42.7 | 19.47 | 0.12 | 19.78 | 0.16 | 19.83 | 0.12 |
| MM47 | | | 3 | 44 | 39.99 | 32 | 08 | 40.5 | 18.27 | 0.04 | 18.63 | 0.06 | 19.19 | 0.08 |
| MM48 | | | 3 | 44 | 40.00 | 32 | 08 | 55.7 | 19.27 | 0.10 | 19.29 | 0.10 | 21.29 | 0.32 |
| MM49 | | | 3 | 44 | 42.88 | 32 | 08 | 48.3 | 19.60 | 0.11 | 19.42 | 0.13 | 20.00 | 0.16 |
| MM5 | NTC013-02 | | 3 | 44 | 31.82 | 32 | 09 | 44.0 | 16.64 | 0.02 | 16.73 | 0.02 | 17.04 | 0.02 |
| MM50 | | | 3 | 44 | 43.27 | 32 | 08 | 58.7 | 19.63 | 0.16 | 20.70 | 0.34 | 20.54 | 0.20 |
| MM51 | | | 3 | 44 | 45.01 | 32 | 08 | 35.5 | 19.74 | 0.17 | 19.46 | 0.11 | 19.51 | 0.11 |
| MM52 | | | 3 | 44 | 45.50 | 32 | 08 | 40.2 | 19.49 | 0.12 | 20.00 | 0.17 | 20.10 | 0.15 |
| MM53 | | | 3 | 44 | 34.10 | 32 | 08 | 37.3 | 17.75 | 0.03 | 18.07 | 0.03 | 18.42 | 0.04 |
| MM55 | | | 3 | 44 | 21.03 | 32 | 08 | 58.6 | 18.51 | 0.05 | 18.89 | 0.06 | 19.07 | 0.07 |
| MM56 | | | 3 | 44 | 27.32 | 32 | 09 | 33.5 | 19.40 | 0.09 | 19.35 | 0.09 | 19.96 | 0.11 |
| MM57 | | | 3 | 44 | 20.11 | 32 | 08 | 49.6 | 18.72 | 0.08 | 19.28 | 0.13 | 19.19 | 0.08 |
| MM58 | | | 3 | 44 | 19.68 | 32 | 08 | 48.6 | 18.76 | 0.08 | 19.21 | 0.11 | 19.17 | 0.08 |
| MM59 | | | 3 | 44 | 33.13 | 32 | 09 | 22.7 | 19.00 | 0.08 | 19.64 | 0.10 | 19.43 | 0.09 |
| MM6 | | | 3 | 44 | 28.91 | 32 | 09 | 39.5 | 18.27 | 0.03 | 18.39 | 0.04 | 18.92 | 0.05 |
| MM60 | | | 3 | 44 | 28.42 | 32 | 10 | 29.2 | 19.60 | 0.12 | 19.15 | 0.08 | 20.07 | 0.16 |
| MM61 | | | 3 | 44 | 32.00 | 32 | 09 | 13.8 | 18.83 | 0.09 | 18.81 | 0.07 | 18.64 | 0.07 |
| MM62** | | | 3 | 44 | 30.79 | 32 | 09 | 04.8 | 19.28 | 0.12 | 20.19 | 0.24 | 20.57 | 0.27 |
| MM63 | | | 3 | 44 | 30.29 | 32 | 09 | 04.5 | 20.22 | 0.20 | 19.73 | 0.14 | 21.03 | 0.31 |
| MM64 | | | 3 | 44 | 31.12 | 32 | 10 | 39.9 | 19.31 | 0.11 | 19.32 | 0.13 | 19.69 | 0.12 |
| MM66 | | | 3 | 44 | 32.23 | 32 | 09 | 55.9 | 18.39 | 0.05 | 18.57 | 0.05 | 19.02 | 0.07 |
| MM67 | | | 3 | 44 | 28.58 | 32 | 10 | 04.0 | 20.25 | 0.19 | 20.61 | 0.19 | 21.06 | 0.30 |
| MM68 | | | 3 | 44 | 29.52 | 32 | 09 | 07.4 | 19.47 | 0.11 | 19.46 | 0.11 | 20.06 | 0.14 |
| MM69 | | | 3 | 44 | 29.80 | 32 | 08 | 58.4 | 19.50 | 0.12 | 19.54 | 0.10 | 20.04 | 0.13 |
| MM7 | NTC095-02 | | 3 | 44 | 29.48 | 32 | 10 | 25.9 | 17.04 | 0.02 | 17.11 | 0.02 | 17.40 | 0.02 |
| MM70 | | | 3 | 44 | 34.19 | 32 | 09 | 14.7 | 19.13 | 0.10 | 19.24 | 0.11 | 19.45 | 0.12 |
| MM71 | | | 3 | 44 | 34.21 | 32 | 09 | 04.3 | 19.60 | 0.16 | 19.38 | 0.11 | 20.18 | 0.18 |
| MM72 | | | 3 | 44 | 31.46 | 32 | 11 | 02.9 | 19.31 | 0.09 | 19.47 | 0.12 | 19.98 | 0.15 |
| MM73 | | | 3 | 44 | 39.72 | 32 | 10 | 28.2 | 20.46 | 0.23 | 20.11 | 0.21 | 21.90 | 1.21 |
| MM74 | | | 3 | 44 | 42.52 | 32 | 10 | 08.6 | 19.98 | 0.21 | 19.83 | 0.18 | 21.68 | 0.61 |
| MM75 | | | 3 | 44 | 39.16 | 32 | 10 | 24.6 | 19.97 | 0.19 | 19.85 | 0.16 | 20.85 | 0.27 |
| MM76 | | | 3 | 44 | 40.52 | 32 | 09 | 43.1 | 19.14 | 0.09 | 18.92 | 0.08 | 19.59 | 0.12 |
| MM78 | | | 3 | 44 | 41.86 | 32 | 09 | 20.4 | 20.14 | 0.21 | 20.46 | 0.28 | 21.13 | 0.30 |
| MM79 | | | 3 | 44 | 35.35 | 32 | 08 | 36.9 | 18.72 | 0.07 | 18.83 | 0.07 | 19.61 | 0.11 |
| MM8 | NTC095-01 | | 3 | 44 | 29.05 | 32 | 10 | 32.1 | 18.12 | 0.05 | 18.32 | 0.05 | 18.76 | 0.05 |
| MM80 | | | 3 | 44 | 49.18 | 32 | 08 | 44.3 | 19.18 | 0.11 | 19.50 | 0.13 | 19.58 | 0.12 |
| MM81 | | | 3 | 44 | 47.44 | 32 | 08 | 33.6 | 17.63 | 0.02 | 17.87 | 0.03 | 18.01 | 0.03 |
| MM82 | | | 3 | 44 | 46.24 | 32 | 08 | 27.2 | 18.28 | 0.04 | 18.20 | 0.04 | 18.73 | 0.06 |
| MM83 | | | 3 | 44 | 37.75 | 32 | 08 | 31.3 | 17.17 | 0.02 | 17.17 | 0.02 | 17.67 | 0.02 |
| MM84 | | | 3 | 44 | 33.96 | 32 | 08 | 40.9 | 18.64 | 0.08 | 19.33 | 0.13 | 19.27 | 0.10 |
| MM85 | | | 3 | 44 | 31.63 | 32 | 08 | 55.4 | 19.64 | 0.20 | 19.97 | 0.25 | 20.34 | 0.25 |
| MM86 | | | 3 | 44 | 35.66 | 32 | 10 | 49.3 | 18.77 | 0.07 | 19.13 | 0.08 | 19.29 | 0.08 |
| MM87 | | | 3 | 44 | 26.43 | 32 | 10 | 31.1 | 20.44 | 0.25 | 19.64 | 0.13 | 20.39 | 0.19 |
| MM88 | | | 3 | 44 | 37.81 | 32 | 11 | 12.6 | 18.45 | 0.06 | 18.71 | 0.06 | 19.02 | 0.09 |
| MM89 | | | 3 | 44 | 42.65 | 32 | 10 | 28.4 | 18.72 | 0.08 | 18.75 | 0.08 | 19.12 | 0.08 |
| MM9 | NTC072-04 | | 3 | 44 | 36.03 | 32 | 11 | 00.5 | 17.98 | 0.03 | 17.97 | 0.03 | 18.34 | 0.03 |
| MM90 | | | 3 | 44 | 45.13 | 32 | 10 | 28.9 | 19.67 | 0.17 | 20.11 | 0.34 | 20.51 | 0.26 |
| MM91 | | | 3 | 44 | 43.87 | 32 | 10 | 00.6 | 19.61 | 0.13 | 19.71 | 0.16 | 20.49 | 0.20 |
| MM92 | | | 3 | 44 | 42.86 | 32 | 11 | 15.9 | 18.45 | 0.13 | 18.63 | 0.20 | 19.05 | 0.19 |
| MM93** | | | 3 | 44 | 39.21 | 32 | 11 | 33.4 | 17.84 | 0.07 | 18.00 | 0.10 | 18.91 | 0.14 |
| MM94 | | | 3 | 44 | 34.38 | 32 | 11 | 21.1 | 20.16 | 0.22 | 19.32 | 0.13 | 19.87 | 0.15 |
| MM95 | | | 3 | 44 | 28.65 | 32 | 11 | 11.0 | 19.26 | 0.11 | 19.01 | 0.09 | 19.65 | 0.12 |
| MM96 | | | 3 | 44 | 35.78 | 32 | 11 | 34.7 | 19.49 | 0.17 | 19.39 | 0.20 | 20.30 | 0.34 |
| MM97 | | | 3 | 44 | 37.12 | 32 | 11 | 37.1 | 19.52 | 0.23 | 19.01 | 0.19 | 19.43 | 0.18 |
| MM98 | | | 3 | 44 | 34.55 | 32 | 10 | 33.5 | 19.93 | 0.15 | 19.33 | 0.13 | 22.60 | 0.95 |
| MM99 | | | 3 | 44 | 31.82 | 32 | 10 | 59.6 | 20.94 | 0.28 | 20.90 | 0.26 | 20.83 | 0.21 |
| MM100 | | | 3 | 44 | 29.35 | 32 | 11 | 03.0 | 19.80 | 0.14 | 20.26 | 0.23 | 20.30 | 0.14 |
| MM101 | | | 3 | 44 | 38.39 | 32 | 11 | 21.9 | 20.16 | 0.33 | 19.58 | 0.19 | 20.51 | 0.33 |
| MM102 | | | 3 | 44 | 34.95 | 32 | 09 | 03.5 | 19.69 | 0.15 | 20.48 | 0.26 | 20.11 | 0.12 |
| MM104** | | | 3 | 44 | 37.37 | 32 | 08 | 39.6 | 19.40 | 0.14 | 19.93 | 0.21 | 20.89 | 0.37 |
| MM105 | | | 3 | 44 | 42.28 | 32 | 08 | 45.4 | 19.54 | 0.16 | 19.77 | 0.12 | 23.65 | 3.45 |
| MM106 | | | 3 | 44 | 43.41 | 32 | 09 | 00.8 | 19.44 | 0.11 | 19.91 | 0.18 | 20.61 | 0.21 |
| MM107 | | | 3 | 44 | 39.29 | 32 | 09 | 32.1 | 19.81 | 0.20 | 20.19 | 0.24 | 20.99 | 0.29 |
| MM108 | | | 3 | 44 | 38.15 | 32 | 09 | 16.3 | 19.66 | 0.12 | 19.60 | 0.14 | 19.48 | 0.11 |
| MM109 | | | 3 | 44 | 39.57 | 32 | 09 | 04.3 | 20.69 | 0.59 | 21.03 | 0.75 | 21.84 | 0.63 |
| MM110 | | | 3 | 44 | 37.61 | 32 | 08 | 49.7 | 19.73 | 0.18 | 19.81 | 0.17 | 20.50 | 0.21 |
| MM111 | | | 3 | 44 | 42.37 | 32 | 08 | 33.9 | 20.50 | 0.30 | 21.88 | 0.72 | 20.50 | 0.30 |
| MM113 | | | 3 | 44 | 28.06 | 32 | 08 | 20.4 | 18.55 | 0.08 | 18.44 | 0.06 | 19.43 | 0.14 |
| MM114 | | | 3 | 44 | 28.06 | 32 | 08 | 26.1 | 20.11 | 0.22 | 20.09 | 0.19 | 20.45 | 0.23 |
| MM115 | | | 3 | 44 | 28.37 | 32 | 08 | 23.6 | 19.55 | 0.16 | 19.41 | 0.13 | 20.56 | 0.28 |
| MM116 | | | 3 | 44 | 23.71 | 32 | 08 | 14.5 | 17.66 | 0.03 | 17.81 | 0.04 | 17.98 | 0.04 |
| MM117 | | | 3 | 44 | 26.54 | 32 | 08 | 10.0 | 16.93 | 0.02 | 16.99 | 0.02 | 17.60 | 0.03 |
| MM118 | | | 3 | 44 | 49.99 | 32 | 08 | 42.4 | 17.97 | 0.06 | 18.36 | 0.09 | 18.55 | 0.10 |
| MM119 | | | 3 | 44 | 24.24 | 32 | 10 | 45.7 | 19.54 | 0.16 | 18.87 | 0.09 | 20.32 | 0.28 |
| MM120 | | | 3 | 44 | 48.10 | 32 | 09 | 49.9 | 18.01 | 0.04 | 18.51 | 0.07 | 18.61 | 0.05 |
| MM121 | | | 3 | 44 | 49.49 | 32 | 09 | 47.3 | 17.88 | 0.05 | 17.82 | 0.07 | 18.45 | 0.08 |
| MM122 | | | 3 | 44 | 49.89 | 32 | 09 | 39.3 | 18.43 | 0.08 | 18.57 | 0.13 | 18.99 | 0.13 |
| MM123 | | | 3 | 44 | 20.75 | 32 | 09 | 31.7 | 19.15 | 0.08 | 19.25 | 0.10 | 20.41 | 0.17 |
| MM124** | | | 3 | 44 | 19.76 | 32 | 09 | 12.3 | 19.42 | 0.12 | 19.62 | 0.13 | 21.41 | 0.44 |
| MM125 | | | 3 | 44 | 19.21 | 32 | 09 | 17.1 | 18.96 | 0.09 | 18.98 | 0.11 | 19.39 | 0.12 |
| MM126** | | | 3 | 44 | 24.80 | 32 | 09 | 55.6 | 19.35 | 0.11 | 19.83 | 0.18 | 20.50 | 0.30 |
| MM127 | | | 3 | 44 | 25.28 | 32 | 10 | 02.1 | 19.50 | 0.13 | 19.83 | 0.17 | 22.42 | 3.41 |
| MM128 | | | 3 | 44 | 28.12 | 32 | 10 | 51.6 | 12.17 | 0.02 | 12.27 | 0.02 | 12.67 | 0.02 |
| MM129 | | | 3 | 44 | 31.18 | 32 | 11 | 25.2 | 16.56 | 0.02 | 16.67 | 0.02 | 16.97 | 0.02 |
| MM130 | | | 3 | 44 | 26.03 | 32 | 10 | 58.9 | 16.04 | 0.02 | 16.08 | 0.02 | 16.57 | 0.02 |
| MM131 | | | 3 | 44 | 26.13 | 32 | 11 | 05.7 | 14.75 | 0.02 | 14.88 | 0.02 | 15.23 | 0.02 |
| MM132 | | | 3 | 44 | 21.10 | 32 | 10 | 04.8 | 17.00 | 0.03 | 17.12 | 0.03 | 17.59 | 0.03 |
| MM133 | | | 3 | 44 | 19.34 | 32 | 08 | 49.3 | 18.75 | 0.08 | 18.95 | 0.07 | 19.39 | 0.11 |
| MM134 | | | 3 | 44 | 18.95 | 32 | 08 | 33.9 | 16.68 | 0.02 | 16.86 | 0.02 | 17.13 | 0.02 |
| MM135 | | | 3 | 44 | 18.26 | 32 | 08 | 33.9 | 16.25 | 0.02 | 16.49 | 0.02 | 16.71 | 0.02 |
| MM136 | | | 3 | 44 | 17.54 | 32 | 08 | 35.8 | 16.24 | 0.02 | 16.49 | 0.03 | 16.69 | 0.02 |
| MM137 | | | 3 | 44 | 31.54 | 32 | 08 | 10.8 | 16.70 | 0.02 | 16.78 | 0.02 | 17.20 | 0.02 |
| MM138 | | | 3 | 44 | 44.62 | 32 | 08 | 11.1 | 11.58 | 0.02 | 11.63 | 0.02 | 12.03 | 0.02 |
| MM139 | | | 3 | 44 | 45.18 | 32 | 08 | 15.7 | 17.45 | 0.03 | 17.61 | 0.03 | 17.66 | 0.03 |
| MM140 | | | 3 | 44 | 45.08 | 32 | 08 | 10.3 | 17.34 | 0.06 | 17.54 | 0.08 | 17.61 | 0.06 |
| MM141 | | | 3 | 44 | 45.77 | 32 | 08 | 10.3 | 18.64 | 0.11 | 18.62 | 0.13 | 20.01 | 0.28 |
| MM142 | | | 3 | 44 | 46.23 | 32 | 08 | 08.5 | 16.13 | 0.02 | 16.31 | 0.02 | 16.48 | 0.02 |
| MM143 | | | 3 | 44 | 41.42 | 32 | 09 | 10.7 | 20.53 | 0.24 | 20.25 | 0.24 | 20.36 | 0.21 |
| MM144 | | | 3 | 44 | 45.01 | 32 | 10 | 26.4 | 19.06 | 0.10 | 18.96 | 0.09 | 19.20 | 0.08 |
| MM145 | | | 3 | 44 | 26.11 | 32 | 08 | 04.6 | 18.34 | 0.07 | 18.48 | 0.09 | 18.92 | 0.10 |
| MM146 | | | 3 | 44 | 42.71 | 32 | 09 | 44.8 | 20.45 | 0.26 | 19.93 | 0.15 | 22.20 | 0.35 |
| MM147 | | | 3 | 44 | 39.64 | 32 | 09 | 56.2 | 20.07 | 0.25 | 20.41 | 0.31 | 20.36 | 0.20 |
| MM149 | | | 3 | 44 | 36.06 | 32 | 08 | 57.7 | 20.37 | 0.42 | 20.17 | 0.27 | 22.01 | 1.22 |

Table 4: IC348 sources detected by FLITECAM. A * indicates that the source was identified in another survey; "NTC" = Najita, Tiede, & Carr (2000); "H" = Herbig (1998); "LRLL" = Luhman, Rieke, Lada, & Lada (1998). New objects detected by our survey are identified by MM. Sources marked with ** are considered potential M, L, and T dwarf candidates based on their $H_2O$ and $CH_4$ magnitudes. RA and Dec are given in J2000 coordinates.